\documentclass[12pt]{aastex7}

\hypersetup{linkcolor=red,citecolor=blue,filecolor=cyan,urlcolor=magenta}
\usepackage{verbatim}  

\newcommand{\Lya}     {Ly$\alpha$}    
\newcommand{\Lyb}     {Ly$\beta$}      

\newcommand {\HI}        {\ion{H}{1}}   
\newcommand {\HII}       {\ion{H}{2}}   

\newcommand {\HeI}     {\ion{He}{1}}   
\newcommand {\HeII}     {\ion{He}{2}}  

\newcommand {\CII}     {\ion{C}{2}}   

\newcommand {\MgI}      {\ion{Mg}{1}}  
\newcommand {\MgII}     {\ion{Mg}{2}}

\newcommand {\SII}        {\ion{S}{2}}

\newcommand {\SiIII}  {\ion{Si}{3}}

\newcommand{\NeVII}    {\ion{Ne}{7}}      
\newcommand {\NeVIII}  {\ion{Ne}{8}}       

\newcommand {\FeVIII}    {\ion{Fe}{8}}
\newcommand {\FeIX}      {\ion{Fe}{9}}
\newcommand {\FeX}       {\ion{Fe}{10}}

\newcommand {\kms}    {km~s$^{-1}$}

\newcommand {\etal}   {et~al.}

\shorttitle{Local-Cloud Photoionization Sources}
\shortauthors{Shull et al.}

\begin{document}

\title{Ionization Sources of the Local Interstellar Clouds:  \\
Two B-stars, Three White Dwarfs, and the Local Hot Bubble  } 

\author[0000-0002-4594-9936] {J. Michael Shull} \email{michael.shull@colorado.edu}
\affiliation{Department of Astrophysical \& Planetary Sciences, University of Colorado, Boulder, CO 80309, USA}
\affiliation{Department of Physics and Astronomy, University of North Carolina, Chapel Hill, NC 27599, USA }

\author{Rachel M. Curran} \email{curran@unc.edu}
\affiliation{Department of Physics \& Astronomy, University of North Carolina, Chapel Hill, NC 27599 USA}

\author[0000-0001-8426-1141] {Michael W. Topping} \email{michaeltopping@arizona.edu}
\affiliation{Steward Observatory, University of Arizona, Tucson, AZ 85721, USA}

\author[0000-0002-7597-6935] {Jonathan D. Slavin} \email{slavin@cfa.harvard.edu} 
\affiliation{Center for Astrophysics, Harvard \& Smithsonian, 60 Garden Sreet, Cambridge, MA 02138, USA} 



\begin{abstract}

The dominant sources of photoionizing radiation in the extreme ultraviolet (EUV) incident on 
the exterior of the local interstellar clouds include two nearby early B-type stars, 
$\epsilon$~CMa ($124\pm2$~pc) and $\beta$~CMa ($151\pm5$~pc), three hot dwarfs,
and the local hot bubble (LHB).  Line emission (170--912~\AA) from highly ionized metals 
(Fe, Ne, Mg) in million-degree LHB plasma may be responsible for the elevated ionization 
fractions of helium ($n_{\rm HeII}/n_{\rm He} \approx 0.4$) compared to hydrogen 
($n_{\rm HII} / n_{\rm H} \approx 0.2$) in the local clouds.  We update the stellar parameters 
and ionizing flux for $\beta$~CMa, after correcting the EUV spectra for intervening \HI\ column 
density, $N_{\rm HI} = 1.9\pm0.1\times10^{18}~{\rm cm}^{-2}$, and its hotter effective 
temperature, $T_{\rm eff} \approx 25,000$~K vs.\ 21,000~K for $\epsilon$~CMa. 
These two stars produce a combined H-ionizing photon flux 
$\Phi_{\rm H} \approx 6800\pm1400$~cm$^{-2}~{\rm s}^{-1}$ at the external surface of 
the local clouds.  The hot bubble could produce comparable fluxes, 
$\Phi_{\rm H} =$ 2000--9000~cm$^{-2}$~s$^{-1}$, depending on the amount of metal depletion 
into dust grains that survive sputtering.  The radial velocities and proper motions of $\beta$~CMa 
and $\epsilon$~CMa indicate that both stars passed within $10\pm1$~pc of the Sun 
$4.4\pm0.1$~Myr ago, with 100--200 times higher local ionizing fluxes.   At that time, the local 
clouds were likely farther from the Sun, owing to their transverse motion.  Over the last few Myr, 
EUV radiation from these two stars left a wake of highly ionized gas in a hot, low-density cavity 
produced by past supernova explosions in the Sco-Cen OB association and connected with the 
LHB. 

\end{abstract}


\section{\bf Introduction}  


\bigskip

The interstellar medium (ISM) within 10--15 pc of the Sun has been characterized as a complex 
of 15 diffuse, low-density clouds (S.\ Redfield \& J. L.\ Linsky 2008; J. L.\ Linsky \etal\ 2019) with 
temperatures $T \approx 7500\pm1300$~K and high ionization fractions 
(J. D.\ Slavin \& P. C.\ Frisch 2008) of hydrogen (20\%) and helium (40\%).   Previous reviews of the 
local ISM (P. C.\ Frisch \etal\ 2011;  J. L.\ Linsky \etal\ 2022b) noted the importance of understanding 
these ionization processes, as the Sun and its outer heliosphere encounter the ISM at a velocity of 
26~\kms.  An alternate geometric model for the local ISM was proposed by 
C.\ Gry \& E. B.\ Jenkins (2014),  with a 
single continuous cloud, nonrigid flows, internal shocks, and gradients in velocity and metal depletions.  
A follow-up analysis (S.\ Redfield \& J. L.\ Linsky 2015) argued that additional sight-line data were better 
represented by the multiple cloud model and suggested that strong interstellar magnetic fields 
could provide more cloud rigidity.  The geometry, ionization state, and volume filling factor of the local
clouds remain uncertain, with their dimensions dependent on the adopted hydrogen density.  If we adopt 
 a neutral hydrogen density $n_{\rm HI} \approx 0.2$~cm$^{-3}$  (J. D.\ Slavin \& P. C.\ Frisch 2008;  
 C.\ Gry \&  E. B.\ Jenkins 2017;  P.\ Swaczyna \etal\ 2022), the typical column density 
 $N_{\rm HI} \approx 2\times10^{18}~{\rm cm}^{-2}$  through the clouds suggests an absorber pathlength 
 $L_{\rm HI} = N_{\rm HI}/n_{\rm HI} \approx 3$~pc, leaving considerable space between the clouds.   
 On the other hand, if we adopt the lower value, $n_{\rm HI} = 0.1$~cm$^{-3}$ estimated from \HI\ 
 absorption toward nearby stars  (S.\ Redfield \& J. L.\ Linsky 2008; J.~L.\  Linsky \etal\ 2022a), the 
 discrete clouds would completely fill space out to 4 pc and most of the space out to 10~pc.  
 
\medskip 

In addition to geometrical uncertainties, another puzzle concerns the source and spectra of ionizing 
radiation incident on the external surface of local diffuse clouds.  
The range of \HI\ column densities toward nearby stars and the different ionization fractions of 
H$^+$ and He$^+$ suggest inhomogeneous cloud structures and multiple sources of ionizing radiation. 
Understanding the cloud geometry and ionization rates are the primary motivation for our current study.  
Based on observations  (J.\ Dupuis \etal\ 1995; J.\ Vallerga \& B.\ Welsh 1995, N.\ Craig \etal\ 1997)
by the Extreme Ultraviolet Explorer (EUVE), the stellar contributions to the photoionization 
of the local ISM (J.\ Vallerga 1998) include two early B-type stars ($\epsilon$~CMa and $\beta$~CMa) 
and three hot white dwarfs (G191-B2B, Feige~24, HZ~43A).  In addition to photoionizing radiation from 
hot stars, the structure of the interstellar gas within 100~pc was influenced by explosions of supernovae
(SNe) from the nearby Scorpius--Centaurus OB association (J.\ Ma\'iz-Apell\'aniz 2001; 
D.\ Breitschwerdt \etal\ 1996; B.\ Fuchs \etal\ 2006; C.\ Zucker \etal\ 2025).  The local hot bubble (LHB) 
of million-degree plasma surrounds the local clouds and can produce a significant flux of ionizing photons 
(J.\ Slavin \& P.\ Frisch 2002)  from emission lines of highly ionized Fe, Mg, Si, and Ne.

\medskip

In a previous paper (J. M.\ Shull \etal\ 2025), we updated the stellar parameters of $\epsilon$~CMa,
based on its revised parallax distance, angular diameter, integrated flux, and effective temperature.
Using non-LTE model atmospheres, we derived the intervening \HI\ column density and the \HI\ 
photoionization rate incident on the external surface of the local clouds.  In this paper, we make 
similar calculations of the stellar parameters and ISM column density for $\beta$~CMa.  
After correcting for EUV flux attenuation, we estimate the photoionization rates of \HI\ and \HeI.   
We then construct multiple-source models of the total photoionization rates of the local clouds, 
using the combined EUV flux from five hot stars and the LHB plasma.   The stars include two B-type 
stars ($\epsilon$~CMa and $\beta$~CMa) and three nearby hot white dwarfs.  We also explore 
non-equilibrium ionization effects arising from stellar motions.  The radial velocities and proper 
motions of the two CMa stars indicate that both passed within $10\pm1$~pc of the Sun 
$4.4 \pm0.1$~Myr ago, with local ionizing fluxes 100--200 times larger than today.  

\medskip

In Section~2, we update a consistent set of stellar parameters (radius, mass, effective temperature, 
luminosity) for $\beta$~CMa, based on measurements of its parallax distance, stellar angular diameter,
and total flux.   We employ non-LTE, line-blanketed model atmospheres to anchor the relative fluxes in 
the stellar EUV continuum shortward of the ionization edges of  \HI\ ($\lambda \leq 911.75$~\AA) and 
\HeI\ ($\lambda \leq 504.26$~\AA).  We use the observed fluxes in the EUV (504--720~\AA) and far-UV 
(1000--1400~\AA) to determine the attenuation of the LyC radiation in the stellar atmosphere and derive
the interstellar \HI\ column density, $N_{\rm HI} = (1.9\pm0.1)\times10^{18}~{\rm cm}^{-2}$.  We find that 
$\epsilon$~CMa and $\beta$~CMa have comparable photoionization rates of \HI\ at the outer surface of 
the local clouds.  However, $\beta$~CMa may have a much larger ionization rate of \HeI, owing to its 
higher effective temperature $T_{\rm eff} \approx 25,000$~K, compared to 21,000~K for $\epsilon$~CMa. 
In Section~3, we describe multi-source models for the H and He ionization structure of the local interstellar 
clouds and the low-density ``interstellar tunnel" in their direction. We also model the EUV line emission 
from the LHB, especially the helium-ionizing continuum shortward of 504~\AA.  
In Section 4, we summarize our results and their implications for the local clouds and the ionization 
state of the gas entering  the Sun's heliosphere.

\bigskip

\section{\bf Revised Properties of Beta Canis Majoris}

\bigskip

\subsection{Basic Stellar Parameters} 

\bigskip

The B-type giant star Beta~Canis Majoris ($\beta$~CMa), also known as HD~44743 and Mirzam, 
is one of several bright sources of extreme ultraviolet (EUV) radiation.  Both $\epsilon$~CMa and 
$\beta$~CMa  are located in the direction ($\ell = 232^{\circ}\pm5^{\circ}$, $b = -15^{\circ}\pm4^{\circ})$ 
of a low-density, highly ionized tunnel in the ISM  (P. C.\ Frisch \& D. G.\ York 1983; C.\ Gry \etal\ 1985; 
B. Y.\ Welsh 1991; J. V.\ Vallerga 1998; D. M.\  Sfeir \etal\ 1999; J. L.\ Linsky \etal\ 2019).  
Both stars contribute to the \HI\ photoionization rate of the Local Interstellar Cloud or LIC  
(J.\ Vallerga 1998).  A portion of the $\beta$~CMa ionizing spectrum was measured by EUVE 
between 504--720~\AA, but we have no direct measurements between 720--912~\AA.  

\medskip

In a survey of southern B-type stars, J. R.\ Lesh (1968) listed the spectral type (SpT) of $\beta$~CMa 
as B1~II-III, with apparent and absolute visual magnitudes $m_V = 1.97$~mag, $M_V = -4.57$~mag, 
and a spectrophotometric distance of 203~pc.  (Hereafter, all values of $m_V$ and $M_V$ are 
in magnitudes.)  The SpT of B1~II-III was confirmed by I.\ Negueruela \etal\ (2024). 
The star was too bright for {\it Gaia} to make parallax measurements.
However, the {\it Hipparcos} parallax measurement of $6.62\pm0.22$~mas (F.\ van Leeuwen 2007) 
provided a shorter distance $d = 151\pm5$~pc,  with a distance modulus
($m_V$--$M_V) = 5.895 \pm0.071$, smaller by 0.643~mag compared to 203~pc .  
In other papers, the distance to $\beta$~CMa was quoted variously as 213~pc (R. C.\ Bohlin 1975), 
206~pc (B. D.\ Savage \etal\ 1977; R. C.\ Bohlin \etal\ 1978; J. P.\  Cassinelli \etal\ 1996), and 203~pc 
(B. Y.\ Welsh 1991; J. M.\ Shull \&  M.\ Van Steenberg 1985).  The new distance modulus implies 
an absolute magnitude $M_V = -3.925 \pm 0.071$, similar to the value of $-4.0$ given in 
J.\ Lesh (1968) for B1~III spectral type.  

\medskip

A key physical measurement for $\beta$~CMa is its angular diameter $\theta_d = 0.52\pm0.03$~mas 
(R.\ Hanbury Brown \etal\ 1974) found with the stellar interferometer at the Narrabri Observatory.  
From this and a parallax distance $d = 151\pm5$~pc, we derive a stellar radius of
\begin{equation}
     R = \left( \frac {\theta_d \, d} {2} \right) = 5.87 \times10^{11}~{\rm cm}  
           ~~ (8.44\pm0.56~R_{\odot})   \; , 
\end{equation}  
where we combined the relative errors on $\theta_d$ (5.8\%) and $d$ (3.3\%) in quadrature.  
The revision in the $\beta$~CMa distance from 206~pc to 151~pc also changes its inferred
mass and surface gravity.   L.\ Fossati \etal\ (2015) used the new distance $d = 151\pm5$~pc 
and evaluated the radius and mass from two sets of evolutionary models:  
$R = 7.4^{+0.8}_{-0.9}~R_{\odot}$ and $M = 12.0^{+0.3}_{-0.7}~M_{\odot}$ (tracks from 
C. G.\ Georgy \etal\ 2013) and $R = 8.2^{+0.6}_{-0.5}~R_{\odot}$ and 
$M = 12.6^{+0.4}_{-0.5}~M_{\odot}$ (tracks from L. L.\ Brott \etal\ 2011).  These radii are in good
agreement with the radius $8.44\pm0.56~R_{\odot}$ and the $13\pm1~M_{\odot}$ gravitational 
and evolutionary masses that we derive below. 

\medskip

The absolute bolometric magnitude of $\beta$~CMa also needs revision based on its new distance.
The stellar classification has remained consistent at B1~II-III  (J. R.\ Lesh 1968; L.\ Fossati \etal\ 2015; 
I.\ Negueruela \etal\ 2024).  At this SpT, the absolute magnitude is $M_V = -4.0$ (J. R.\ Lesh 1968 at 
B1.5~III).  We adopt a bolometric correction $(M_{\rm bol}$--$M_V) \approx -2.5$ from Figure~5  
of M. G.\ Pedersen \etal\ (2020) at $T_{\rm eff} = 25,000$~K and $\log g$ = 3.5--4.0.  This yields a 
bolometric absolute magnitude $M_{\rm bol} = -6.425\pm0.07$ and luminosity of $10^{4.47}~L_{\odot}$
based on the solar bolometric absolute magnitude $M_{{\rm bol},\odot} = 4.74$.  The main uncertainty
in this luminosity comes from the bolometric correction.   

\bigskip 

Values of $R$, $T_{\rm eff}$, $g$, and $L$ must satisfy the relations $g =  (G M/R^2)$ and 
$L = 4 \pi R^2 \sigma_{\rm SB} T_{\rm eff}^4$, where 
$\sigma_{\rm SB} = 5.6705 \times 10^{-5}$ erg~cm$^{-2}$~s$^{-1}$~K$^{-4}$ is the 
Stefan-Boltzmann constant and $R = (\theta_d \, d/2)$.  Here, $\theta_d = 2.52 \times10^{-9}$~rad 
($0.52\pm0.03$ mas) is the measured stellar angular diameter (R.\ Hanbury Brown \etal\ 1974).  
The integrated stellar flux is $f = (36.2 \pm 4.9) \times10^{-6}~{\rm erg~cm}^{-2}~{\rm s}^{-1}$ 
from A.\ Code \etal\ (1976), who combined ground-based visual and near-infrared photometry 
with ultraviolet observations (1100--3500~\AA) taken by the Orbiting Astronomical Observatory 
(OAO-2).  This flux included an extrapolated value for shorter wavelengths 
$f(\lambda \leq 1100~{\rm \AA}) = (7.5\pm3.8)\times10^{-6}~{\rm erg~cm}^{-2}~{\rm s}^{-1}$.
Even after correcting the observed EUV flux for interstellar absorption, the Lyman continuum 
represents less than 1\% of the bolometric flux.  The constraints from stellar angular diameter 
and total flux give a useful relation for the bolometric effective temperature 
$T_{\rm eff} = (4f / \sigma_{\rm SB} \, \theta_d^2)^{1/4}$,
\begin{equation} 
    T_{\rm eff} = (25,180\pm1120~{\rm K}) \left( \frac {f} {36.2\times10^{-6}} \right)^{1/4} 
             \left( \frac {\theta_d} {0.52~{\rm mas}} \right)^{-1/2}       \;  . 
\end{equation}
The stellar radius and bolometric relation with $T_{\rm eff}$ suggest a total luminosity,
\begin{equation}
   L \approx (25,800\pm3900~L_{\odot}) \left( \frac {T_{\rm eff}}{25,180~{\rm K}} \right)^4 
          \left( \frac {R}{8.44~R_{\odot}} \right)^2  \; .
\end{equation}
With the relative errors on parallax distance (3.3\%) and integrated flux (13.5\%), this luminosity 
$L = 10^{4.41\pm0.06 }~L_{\odot}$ has a 15\% uncertainty, and it is also 15\% lower 
than our estimate $L = 10^{4.47\pm0.07}$  from photometry and bolometric correction.  Both 
luminosities agree within their propagated uncertainties using the relations 
$L = 4 \pi d^2 f = 4 \pi R^2 \sigma_{\rm SB} T_{\rm eff}^4$.   Our values, 
$T_{\rm eff} = 25,180$~K and $L = 10^{4.41\pm0.06}$, are similar to those quoted in
L.\ Fossati \etal\ (2015).  {\bf Table 1} summarizes the stellar parameters from this work 
and previous papers.

\bigskip

\subsection{Stellar Atmospheres and EUV Fluxes}

\bigskip

For $\beta$~CMa, J. P.\ Cassinelli \etal\ (1996) discussed two sets of stellar atmosphere 
parameters ($T_{\rm eff}$ and $\log g$).  Following the method of  A.\ Code \etal\ (1976),
they used the observed total flux and angular diameter to find $T_{\rm eff} = 25,180$~K,
the same as ours (eq.\ [2]) using the same method.  However, their value of surface 
gravity ($\log g = 3.4$) is lower than our value of 3.7.   They also discussed a possible lower 
temperature, $T_{\rm eff} = 23,250$~K with $\log g = 3.5$, that was in better agreement 
with their line-blanketed model atmospheres.  However, they also noted that the higher 
$T_{\rm eff}$ models gave better agreement with the 
observed EUV fluxes\footnote{J. P.\ Cassinelli \etal\ (1995) suggested that backwarming by the 
shocked stellar wind could boost the temperature in the upper atmosphere where the Lyman 
continuum is formed.  A similar non-LTE wind effect was proposed by F.\ Najarro \etal\ (1996), 
involving doppler shifts and velocity-induced changes in density that affect the escape of \HI\ 
and \HeI\ resonance lines and their ground-state populations.  Both mechanisms are sensitive 
to mass-loss rates in the range $\dot{M} \approx$ (1--10)$\times10^{-9}~M_{\odot}~{\rm yr}^{-1}$.}.
 At their larger adopted distance, $d = 206$~pc, the implied stellar parameters 
 ($R \approx 11.5~R_{\odot}$ and $M = gR^2/G \approx 24~M_{\odot}$) are discrepant with
 the derived position of $\beta$~CMa on the H-R diagram.   In a spectroscopic analysis of 
$\beta$~CMa, L.~Fosatti \etal\ (2015) found $T_{\rm eff} = 24,700\pm300$~K and 
 $\log g = 3.78\pm0.08$.  For consistency  with the key observed parameters (distance, total flux, 
 interferometric diameter) we adopt $T_{\rm eff} = 25,180\pm1120$~K and $\log g = 3.70\pm0.08$ 
 ($g = GM/R^2 = 5010\pm500~{\rm cm~s}^{-2}$).  With a measured rotational velocity  
 $V_{\rm rot} \sin i = 20.3\pm7.2$~\kms\ (L.\ Fosatti \etal\ 2015), we can neglect the centrifugal term, 
 $V_{\rm rot}^2/R \approx (7~{\rm cm~s}^{-2})(\sin i)^{-2}$ and infer a gravitational mass
 ($M = gR^2/G$) of  
\begin{equation}
   M = (13 \pm 3~M_{\odot}) \left( \frac {g}{5010~{\rm cm~s}^{-2}} \right) 
        \left( \frac {R} {8.44~R_{\odot}} \right)^2   \; .
\end{equation}
The quoted error on $M$ includes uncertainties in surface gravity and radius, added in quadrature 
with $(\sigma_M/M)^2 = (\sigma_g/g)^2 + (2 \sigma_R/R)^2$.   This gravitational mass is consistent 
with its evolutionary mass on the H-R diagram, $M_{\rm evol} = 13\pm1~M_{\odot}$. {\bf Figure 1} 
shows its location at $\log (L/L_{\odot}) = 4.41$ and $T_{\rm eff} = 25,180$~K, with evolutionary tracks 
from L.\ Brott \etal\ (2011).  This position is consistent with luminosity class~II/III (giant), and it places 
$\beta$~CMa within the $\beta$~Cephei instability strip, consistent with its observed pulsations
ranging from $m_V =$ 1.97 to 2.01 over a 6-hr period (A. Mazumdar \etal\  2006).
This conclusion is confirmed by comparison with the locus of radial and non-radial instability modes 
shown in Figs.\ 2 and 3  of L.\ Deng \& D.~R.\ Xiong (2001).

\bigskip

\subsection{Interstellar Absorption and \HI\ Column Density } 

\bigskip

The EUVE observations of $\beta$~CMa (J. P.\ Cassinelli \etal\ 1996; N.\ Craig \etal\ 1997) showed
detectable flux between 510~\AA\ and 720~\AA, but no flux was detected below the \HeI\ edge 
($\lambda \leq 504.26$~\AA) or \HeII\ edge ($\lambda < 227.84$~\AA).  The lack of observed 
\HeI\ continuum flux is almost certainly the result of the large optical depth at 504~\AA\ arising 
from \HI\ and \HeI\ photoelectric absorption in the intervening ISM.  The two B-type stars, 
$\epsilon$~CMa  and $\beta$~CMa, are the strongest sources of EUV radiation as viewed from 
low-Earth orbit (J. V.\ Vallerga \& B. Y.\ Welsh 1995).  This is primarily a result of their location in a 
low-density cavity or  interstellar tunnel (C. G.\ Gry \etal\ 1985; B. Y.\ Welsh 1991; J. V.\ Vallerga \etal\ 
1998).  In order to find the photoionizing flux incident on the exterior surface 
of the local clouds, we need to correct for EUV flux attenuation by \HI\ and \HeI\ absorption.

\medskip

Previous estimates of the amount of interstellar \HI\  toward $\beta$~CMa were somewhat 
uncertain.  The {\it Copernicus} spectroscopic fits to the interstellar \Lya\ absorption line wings
set upper limits of $N_{\rm HI} < 5\times10^{18}~{\rm cm}^{-2}$ (R. C.\ Bohlin 1975; 
R. C.\ Bohlin \etal\ 1978) and  $N_{\rm HI} < 3\times10^{18}~{\rm cm}^{-2}$  (J. M.\ Shull \&
M. E.\ Van Steenberg 1985). The upper limit of $N_{\rm HI} <  3\times10^{18}~{\rm cm}^{-2}$ 
corresponds to a Lyman-limit (LL) optical depth $\tau_{\rm LL} < 19$  at 912~\AA, but the actual
 \HI\ column density must be smaller, since some EUV flux penetrates the local clouds.  
C. G.\  Gry \etal\ (1985) fitted {\it Copernicus} data in the
\Lyb\ wings to find $N_{\rm HI} =$ (1.6--2.2)$\times 10^{18}~{\rm cm}^{-2}$, after adjusting 
for \HI\ in the stellar atmosphere.  They also estimated a large column density of ionized hydrogen,
$N_{\rm HII} =$ (1.6--2)$\times10^{19}~{\rm cm}^{-2}$, scaled from the \SII\ column density,
$N_{\rm SII} =$ (1.7--2.2)$\times10^{14}~{\rm cm}^{-2}$ assuming a solar S/H abundance.  
The EUVE fluxes (510--720~\AA) detected by 
EUVE were combined with stellar atmosphere models to estimate an interstellar column density 
$N_{\rm HI} =$ (2.0--2.2)$\times10^{18}~{\rm cm}^{-2}$ (J. P.\ Cassinelli \etal\ 1996).  Below, we
make similar calculations, comparing EUVE fluxes to non-LTE atmospheres.  We find a 
slightly smaller value, $N_{\rm HI} = (1.9\pm0.1)\times10^{18}~{\rm cm}^{-2}$, corresponding
to $\tau_{\rm LL} = 12.0\pm0.6$.  The \HeI\ optical depth could be large, depending on the 
amount of neutral helium associated with the ionized gas toward $\beta$~CMa.

\bigskip

In our previous study of $\epsilon$~CMa (J. M. Shull \etal\ 2025) we derived the intervening
\HI\ column density by comparing the observed EUV and FUV continuum fluxes from 
300--1150~\AA\ to non-LTE model atmospheres.  By exploring various attenuation models, 
we found an interstellar column density $N_{\rm HI} = (6\pm1)\times10^{17}$~cm$^{-2}$. 
For $\beta$~CMa, with less complete coverage of the EUV, we compared EUVE and model 
fluxes at four wavelengths (700, 650, 600, 570~\AA) and anchored to the observed fluxes at 
1000--1200~\AA\ from Voyager and International Ultraviolet Explorer (IUE).  The expected 
FUV/EUV continua were based on model atmospheres produced with the code 
\texttt{WM-basic} developed by A. W. A.\ Pauldrach \etal\ (2001)\footnote{This code can be found at 
http://www.usm.uni-muenchen.de/people/adi/Programs/Programs.html}.
We chose this code because of its hydrodynamic solution of expanding atmospheres with 
line blanketing and non-LTE radiative transfer, including its treatment of the continuum and
wind-blanketing from  EUV lines.  Extensive discussion of hot-star atmosphere codes appears 
in papers by D.~J.\ Hillier \& D. L.\ Miller (1998), F.\ Martins \etal\ (2005), and C.\ Leitherer \etal\ (2014).  
We restore the observed continuum to its shape at the stellar surface by multiplying the observed 
flux by $\exp(\tau_{\lambda})$, using optical depths $\tau_{\lambda}$ of photoelectric absorption 
in the ionizing continua of \HI\  ($\lambda \leq 912$~\AA) and \HeI\ ($\lambda \leq 504$~\AA),
\begin{eqnarray}
 \tau_{\rm HI} (\lambda) &\approx& (6.304) \left( \frac {N_{\rm HI} } {10^{18}~{\rm cm}^{-2} } \right) 
           \left( \frac {\lambda} {912~{\rm \AA} } \right)^3   \; , \\
 \tau_{\rm HeI} (\lambda) &\approx& (0.737) \left( \frac {N_{\rm HeI} } {10^{17}~{\rm cm}^{-2} } \right) 
           \left(  \frac {\lambda} {504~{\rm \AA} } \right)^{1.63}   \; .   
\end{eqnarray}
These approximations are based on power-law fits to the photoionization cross sections at 
wavelengths below threshold,
$\sigma_{\rm HI}(\lambda) \approx (6.304\times10^{-18}~{\rm cm}^{2})(\lambda/912~{\rm \AA})^3$
(D. E.\ Osterbrock \& G.\ Ferland 2006) and 
$\sigma_{\rm HeI}(\lambda) \approx (7.37\times10^{-18}~{\rm cm}^{2})(\lambda/504~{\rm \AA})^{1.63}$
fitted to data from D.\ Samson \etal\ (1994).  In our actual calculations, we used the exact 
(non-relativistic) \HI\ cross section (H. A.\ Bethe \& E. E.\ Salpeter 1957),
\begin{equation} 
   \sigma_{\nu} = \sigma_0 \left( \frac {\nu}{\nu_0} \right)^{-4} 
             \frac {\exp[ 4 - (4 \arctan \epsilon ) / \epsilon ]}  { [1 - \exp(-2 \pi / \epsilon) ] }  \; .
\end{equation}
Here, the dimensionless parameter $\epsilon \equiv [(\nu/\nu_0) -1]^{1/2}$ with frequency $\nu_0$ 
defined at the ionization energy $h \nu_0 = 13.598$~eV and 
$\sigma_0 =  6.304\times10^{-18}~{\rm cm}^{2}$.  The two formulae agree at threshold $\nu = \nu_0$, 
but the approximate formula deviates increasingly at shorter wavelengths.  The exact cross section
in eq.\ (7)  is higher than the $\sigma_0 (\nu/\nu_0)^{-3}$ approximation by 8.2\% (700~\AA), 
12.3\% (600~\AA), and 16.4\% (500~\AA), resulting in a lower $N_{\rm HI}$. 
 
 \medskip
   
In the FUV, the stellar continuum flux in the non-LTE model atmosphere rises slowly from 1150~\AA\ 
down to 1000~\AA\ and then declines owing to absorption in higher Lyman-series lines converging 
on the LL at 911.75~\AA.  A pure blackbody at $T = 25,180$~K peaks in $F_{\lambda}$ at 
$\lambda_{\rm max} \approx 1160$~\AA.  However, the spectral shape is uncertain from 1150~\AA\ 
down to 912~\AA, a region unobserved by IUE.  The Voyager observations, plotted in Figure 2a of 
J. P.\ Cassinelli \etal\ (1996), show a peak in 
$F_{\lambda} \approx 4\times10^{-8}~{\rm erg~cm}^{-2}~{\rm s}^{-1}~{\rm \AA}^{-1}$ around 
1000~\AA, similar to the mean flux seen in IUE (SWP Large Aperture) archival spectra at
1150~\AA.    For our flux attenuation calculations, we adopt a continuum level at $912^+$~\AA, just 
longward of the Lyman edge, of 
$F_{\lambda} = 4\times10^{-8}~{\rm erg~cm}^{-2}~{\rm s}^{-1}~{\rm \AA}^{-1}$.
This corresponds to a photon flux of
 $\Phi(912^+) = 1836$ photons~cm$^{-2}~{\rm s}^{-1}~{\rm \AA}^{-1}$.  
 
 \medskip
 
 In the  \texttt{WM-basic} model atmosphere {\bf (Figure 2)}, the stellar flux drops by a factor 
 $\Delta_{\rm star} \equiv F(912^+)/F(912^-) = 67\pm7$ at the Lyman edge.  
We then compare the model fluxes to the photon fluxes $\Phi(\lambda) = \lambda F_{\lambda}/hc$ 
at four EUVE-observed wavelengths, with photon fluxes (photons cm$^{-2}$~s$^{-1}$~\AA$^{-1}$) 
of 0.02 (700~\AA), 0.05 (650~\AA), 0.085 (600~\AA), and 0.11 (570~\AA).  After converting 
$F_{\lambda}$ to $\Phi_{\lambda}$, we predict model photon flux ratios of
 \begin{eqnarray}
    \Delta(700) &=& \Phi(912^+)/\Phi(700) = 209  \\
    \Delta(650) &=& \Phi(912^+)/\Phi(650) = 468  \\
    \Delta(600) &=& \Phi(912^+)/\Phi(600) = 788  \\
    \Delta(570) &=& \Phi(912^+)/\Phi(570) = 1179   \;  .
 \end{eqnarray} 
 Combining the observed photon fluxes, model flux ratios, and anchor flux $\Phi(912^+)$, we 
 arrive at estimates of the ISM optical depth,
 \begin{equation}
    \tau_{\rm ISM} = \ln \left[ \frac {\Phi(912^+)/\Phi(\lambda)} {\Delta(\lambda)} \right]   \; ,
 \end{equation} 
 with a corresponding \HI\ column density $N_{\rm HI} = \tau_{\rm ISM} / \sigma_{\rm HI} (\lambda)$.  
 The most reliable estimates come from data and models at 
 700~\AA\ ($\sigma_{\rm HI} = 3.086\times10^{-18}~{\rm cm}^{2}$) and
 650~\AA\ ($\sigma_{\rm HI} = 2.516\times10^{-18}~{\rm cm}^{2}$), for which we obtain:
 \begin{eqnarray}
     \tau_{\rm ISM}(700~{\rm \AA}) &=&  5.964 \; \; {\rm and} \; \; 
            N_{\rm HI} = 1.93\times10^{18}~{\rm cm}^{-2}   \\
     \tau_{\rm ISM}(650~{\rm \AA}) &=&  4.608 \; \; {\rm and} \; \; 
            N_{\rm HI} = 1.83\times10^{18}~{\rm cm}^{-2}  \;  .
 \end{eqnarray}  
 We adopt $N_{\rm HI} = (1.9\pm0.1)\times10^{18}~{\rm cm}^{-2}$ with optical depth 
 $\tau_{\rm ISM} = 12.0 \pm 0.6$ at the LL corresponding to a  flux decrement factor 
 $\Delta_{\rm ISM} \approx e^{12} \approx 1.6\times 10^5$.  The stellar atmosphere produces 
 a LL decrement of $\Delta_{\rm star} \approx 67\pm7$.  With this much attenuation, there was
 no detectable flux from 750~\AA\ to 912~\AA.  The EUVE spectrometers did not observe 
 at $\lambda > 730$~\AA, and the Colorado Dual-Channel Extreme Ultraviolet Continuum Experiment 
 (DEUCE) spectrograph saw no flux below 912~\AA\ in its (2 November 2020) rocket flight 
 (Emily Witt, private communication).  
 
 \bigskip

\subsection{Photoionization Rates from $\beta$~CMa} 

\bigskip

From our derived interstellar column density toward $\beta$~CMa of
$N_{\rm HI} = (1.9\pm0.1)\times10^{18}~{\rm cm}^{-2}$, we adopt a \HeI\ column density 
$N_{\rm HeI} = (1.3\pm0.2)\times10^{17}~{\rm cm}^{-2}$, based on the mean ratio
$N_{\rm HI} / N_{\rm HeI}  \approx 14$ of interstellar \HI\ and \HeI\ seen toward nearby 
white dwarfs\footnote{J.\ Dupuis \etal\ (1995) reported a mean 
ratio $\beta = N_{\rm HI} / N_{\rm HeI}  \approx 14$, correcting the observed \HI\ and \HeI\ 
column densities for stellar contributions in LTE, hydrostatic, plane-parallel model atmospheres.  
The EUVE  column densities indicated $\beta$ ranging from $12.1\pm3.6$ to $18.4\pm1.3$.  
Removing the two extreme values (the value of 12.1 has 30\% uncertainty) we obtain a mean 
and rms dispersion of $\beta = 14.5\pm1.1$.
These values are consistent with previous sounding-rocket results (J.\ Green \etal\ 1990) with
$11.5 < \beta < 24$ (90\% confidence level) toward the white dwarf G191-B2B.}.  The \HeI\
column density could be somewhat larger if a detectable amount of \HeI\ is associated 
with the estimated column density, $N({\rm H}^+) \approx 10^{19}~{\rm cm}^{-2}$ of ionized 
hydrogen along the sight line (C. G.\ Gry \etal\ 1985' O.\ Dupin \& C. G.\ Gry 1998; 
E. B.\ Jenkins \etal\ 2000).   Much of this H$^+$ may occur near $\beta$~CMa.  From uncertain 
EUVE measurements at the \HeI\ ionization edge,  J. P.\  Cassinelli \etal\ (1996) estimated 
$N_{\rm HeI} \geq 1.4\times10^{18}~{\rm cm}^{-2}$, while J. P.\ Aufdenberg \etal\ (1999) estimated 
$N_{\rm HeI} \geq 6\times10^{17}~{\rm cm}^{-2}$.  However, both these values are inconsistent 
with the LIC ratios of $N_{\rm HI}/N_{\rm HeI} \approx 14$ seen toward six nearby white dwarfs.  

\medskip

We derive the photoionization rates of \HI\ and \HeI\ by integrating the ionizing photon flux
$\Phi_{\lambda} =  (\lambda \, F_{\lambda} / hc)$ multiplied by the photoionization cross section
$\sigma_{\lambda}$ and the attenuation factor $\exp (\tau_{\lambda})$.  
For \HI\ ionization, we use the detected EUVE fluxes (504--730~\AA) and an extrapolation of
the product $\Phi_{\lambda} \sigma_{\lambda} \exp (\tau_{\lambda})$ from 730~\AA\ to 912~\AA.
After correcting for ISM attenuation, we integrate over the ionizing spectrum to find a total photon 
flux of $\Phi_{\rm H} = 3700\pm1000~{\rm cm}^{-2}~{\rm s}^{-1}$ and an \HI\ photoionization rate 
$\Gamma_{\rm HI} = 1.5\times10^{-14}~{\rm s}^{-1}$ at the external surface of the local cloud.  
These values are comparable to those of $\epsilon$~CMa (J. M.\ Shull \etal\ 2025) with 
$\Phi_{\rm H} = 3100\pm1000~{\rm cm}^{-2}~{\rm s}^{-1}$ and 
$\Gamma_{\rm HI} = 1.3\times10^{-14}~{\rm s}^{-1}$.

\medskip

For the \HeI\ continuum, we again use the non-LTE model atmosphere  {\bf (Figure 2)} to establish 
the flux ratio between the two ionization edges, $F(912^+)/F(504^+) \approx 7000$.  The flux falls off 
rapidly below the \HeI\  threshold and is well fitted between $400~{\rm \AA} \leq \lambda \leq 504$~\AA\ 
by the relation $F_{\lambda} = F_0 (\lambda/\lambda_0)^{10}$, with 
$F_0 = 5.7\times10^{-12}~{\rm erg~cm}^{-2}~{\rm s}^{-1}~{\rm \AA}^{-1}$ at 
$\lambda_0 = 504.26$~\AA. The \HeI\ photoionization cross section is
$\sigma_{\rm HeI} \approx \sigma_0 (\lambda/\lambda_0)^{1.63}$, where 
$\sigma_0 = 7.37\times10^{-18}~{\rm cm}^2$.  We integrate over wavelengths $\lambda \leq \lambda_0$, 
defining a dimensionless variable $u = (\lambda / \lambda_0)$ and expressing the surface photoionization 
integral as
\begin{equation}
   \Gamma_{\rm HeI} = \int_{0}^{\lambda_0}  \left[ \frac {\lambda \, F_{\lambda}}
         {hc} \right] \, \sigma_{\rm HeI} (\lambda) \, \exp [\tau(\lambda)]  \, d \lambda  
   = \left[ \frac {F_0 \, \sigma_0 \, \lambda_0^2} {hc} \right] \int_{0}^{1} u^{12.63} \, 
       \exp[ \tau(u)] \,   du \;  .
\end{equation} 
Similarly, the total photon flux in the \HeI\ ionizing continuum can be expressed as,
\begin{equation}
   \Phi_{\rm HeI} = \int_{0}^{\lambda_0}  \left[ \frac {\lambda \, F_{\lambda} }
         {hc} \right] \;  \exp [\tau(\lambda)] \,  d \lambda  
   = \left[ \frac {F_0 \, \lambda_0^2} {hc} \right] \int_{0}^{1} u^{11} \, \exp [\tau(u)] \,    du \;  .
\end{equation} 
The two dimensionless integrals are 1.35 (eq.\ [15]) and 1.46 (eq.\ [16]), showing increases by
factors of 18 (mean $\tau = 2.86$) over the unattenuated integrals (setting $e^{\tau} = 1$).  We 
then obtain $\Gamma_{\rm HeI} = 7.3\times10^{-16}~{\rm s}^{-1}$ and 
$\Phi_{\rm HeI} = 107$ photons cm$^{-2}$~s$^{-1}$.  Because of its hotter effective temperature, 
$\beta$~CMa is a larger source of He-ionizing photons in the local cloud than $\epsilon$~CMa,
for which we estimate $\Gamma_{\rm HeI} = 4.4\times10^{-17}$~s$^{-1}$.   

\newpage 

\section{\bf Photoionization of the Local Clouds}

\bigskip

\subsection{Coupled Hydrogen and Helium Ionization} 
 
 \bigskip
 
In our multi-source models, we assume a cloud of constant total density of hydrogen,
$n_{\rm H} \equiv n_{\rm HI} + n_{\rm HII} = 0.2~{\rm cm}^{-3}$, with $n_{\rm He} = 0.1 n_{\rm H}$.  
The estimated LIC density has varied in the literature, as have the ionization fractions of
hydrogen and helium, defined as $x = n_{\rm HII}/n_{\rm H}$ and $y = n_{\rm HeII}/n_{\rm He}$.
Because of EUV photoelectric absorption in the \HI\ and \HeI\ continua, these fractions
change with depth into the clouds.  Because the absorption cross sections fall off rapidly at
shorter wavelengths below the ionization edges (912~\AA\ for \HI\ and 504~\AA\ for \HeI) the
fractions at the external surface of the clouds differ from those of gas that enters the heliosphere. 
Similarly, the EUV fluxes observed by the EUVE spacecraft are greatly attenuated from those 
outside the local clouds.

\medskip

The key observational constraints on parameters ($n_{\rm H}$, $x$, $y$) are the electron density, 
$n_e = 0.10\pm0.04~{\rm cm}^{-3}$ from models of interstellar species (P. C.\ Frisch \etal\ 2011), the 
elevated ratio of column densities, $N_{\rm HI}/N_{\rm HeI} \approx 14$ (J.\ Dupuis \etal\ 1995) 
measured by EUVE toward local white dwarfs, and the density of \HeI\ neutrals, 
$n_{\rm HeI} = 0.015\pm0.002~{\rm cm}^{-3}$, flowing into the heliosphere (G. Gloeckler \etal\ 2009).   
{\bf Appendix~A} describes the range of parameters allowed by these observations in an analytic model 
of a homogeneous local cloud.  Although the ionization fractions vary with depth into the cloud, 
our model (see eqs.\ A5 and A6) suggests average parameters
$n_{\rm H} = 0.30^{+0.03}_{-0.02}~{\rm cm}^{-3}$,  $x = 0.28^{+0.06}_{-0.07}$, and 
 $y = 0.50^{+0.04}_{-0.05}$.  That hydrogen density is comparable to the value,
$n_{\rm H} = 0.30^{+0.10}_{-0.13}~{\rm cm}^{-3}$, derived for the LIC toward the nearby star
$\alpha$~Leo (C. G.\ Gry \& E. B.\ Jenkins 2017) using UV absorption-line measurements from the
 {\it Hubble Space Telescope} with the high-resolution STIS echelle spectrograph.  
A related estimate of the helium ionization comes from measurements and filtration models of 
interstellar neutrals penetrating the heliosphere of the solar wind.   If we combine the solar 
heliosphere density $n_{\rm HeI} = 0.015~{\rm cm}^{-3}$ with the derived 
$n_{\rm HeII} = 9.7\times10^{-3}~{\rm cm}^{-3}$ (P.\ Swacyzna \etal\ 2023) in the inflowing local 
ISM outside the termination shock, we obtain $n_{\rm He} \approx 0.025~{\rm cm}^{-3}$ and a
helium ionization fraction $y \approx 0.4$.  For a solar helium abundance, 
$n_{\rm He}/n_{\rm H} = 0.1$, this would imply $n_{\rm H} = 0.25~{\rm cm}^{-3}$.

\medskip

Our numerical models follow the spatial variation of H and He ionization and radiative transfer of the 
photoionizing flux distributions with depth into the cloud.  The boundary conditions at the external
cloud surface use the ionizing fluxes and spectra of the five stars and LHB. The stellar EUV fluxes 
at the surface of the local clouds were determined after correcting for attenuation within the cloud
using the inferred column densities $N_{\rm HI}$ for the two B-stars and three white dwarfs.  These 
range from low values, $0.6\times10^{18}~{\rm cm}^{-2}$ ($\epsilon$~CMa) and 
$0.85\times10^{18}~{\rm cm}^{-2}$ (HZ~43A), to higher values (1.5--2.9)$\times10^{18}~{\rm cm}^{-2}$ 
for the other stars.  For the LHB and our illustrative models, we adopt $N_{\rm HI} = 10^{18}~{\rm cm}^{-2}$.  
{\bf Figure 3} shows the positions of the five stars (two B-stars and three white dwarfs) in Galactic coordinates, 
overlaid on the map of background \Lya\ emission (R.\ Gladstone \etal\ 2025) obtained from the New Horizons 
spacecraft on cruise at 57~AU in the outer heliosphere.

\medskip

The two B-type stars, $\epsilon$~CMa and $\beta$~CMa, produce a combined LyC photon flux at the 
external cloud surface of $\Phi_{\rm HI} \approx 6800\pm1400~{\rm cm}^{-2}~{\rm s}^{-1}$ with a hydrogen 
ionization rate $\Gamma_{\rm HI} \approx 3\times10^{-14}~{\rm s}^{-1}$.  Considering photoionization 
equilibrium of hydrogen alone, we can estimate  $x = (1/2)[ -a + (a^2 + 4a)^{1/2}]$, the solution of 
$x^2/(1-x) = a$.  The constant $a = [\Gamma_{\rm HI} / 1.1 n_{\rm H} \alpha_{\rm H}]$, with a factor of 
1.1 correction for helium if its ionization fraction is similar to hydrogen.  We adopt a hydrogen case-B 
radiative recombination coefficient (B. T.\ Draine 2011)
$\alpha_{\rm H} = 3.39\times10^{-13}~{\rm cm}^3~{\rm s}^{-1}$ at $T = 7000$~K to find $a = 0.402$ 
and $x = 0.464$ at the surface of the local clouds.   If we include the low He-ionizing continuum fluxes
(from just $\beta$~CMa and $\epsilon$~CMa) with $\Gamma_{\rm HeI} = 7.3\times10^{-16}$~s$^{-1}$, 
we find $x \approx 0.479$ (H$^+$) and $y \approx 0.022$ (He$^+$) at the cloud surface.  As noted by 
J.\ Dupuis \etal\ (1995) and J. V.\ Vallerga (1998), the three hot white dwarfs (G191-B2B, Feige~24, and 
HZ~43A) are important sources of \HeI\ ionizing flux.  However, when we include them together with
the B-stars, they do not fully explain the elevated level of He$^+$ ionization.  
 
 \bigskip
 
 We are then led to consider ionizing photons from the cavity of hot plasma ($T \approx 10^6$~K) 
 believed responsible for the soft X-ray background (S.L.\ Snowden \etal\ 1990; M.\ Galeazzi \etal\ 2014; 
 M. C. H.\ Yueng \etal\ 2024). 
For more accurate calculations of H and He ionization, we constructed radiative-transfer models 
of the local cloud, including the five major stellar sources of EUV radiation, together with line emission 
from the LHB.  {\bf Figure~4} shows a sample model, with $n_{\rm H} = 0.2~{\rm cm}^{-3}$ and a mean
\HI\ column density $N_{\rm HI} = 10^{18}~{\rm cm}^{-2}$.  We computed the flux attenuation of the 
ionizing spectrum and the accompanying decrease in ionization fractions with depth into the cloud,
based on the integrated optical depths in the \HI\ and \HeI\ ionizing continua.  The attenuated spectra
are calculated with a wavelength grid spacing $\Delta \lambda = 1.58$~\AA\ from 124--912~\AA.
Because the free electrons come from both H$^+$ and He$^+$, we solve coupled equations 
(details in {\bf Appendix B}) for the local H and He ionization fractions ($x$ and $y$) with electron 
fraction $f_e = n_e/n_{\rm H} = (x + 0.1y)$ for a solar abundance ratio $n_{\rm He}/n_{\rm H} = 0.1$.  
As shown in eqs.\ B17 and B18, the local ratio of He and H ionization fractions in photoionization
equilibrium can be expressed as 
 \begin{equation}
   \frac  {y}{x} = \left( \frac {\Gamma_{\rm HeI}} {\Gamma_{\rm HI}} \right) 
         \left[ \frac {1 + (a_{\rm H}/f_e)} {1 + (a_{\rm He}/f_e)} \right] T_{7000}^{-0.020}  \;  .
\end{equation}
This expression shows the dominant importance of the ratio of ionization rates in determining
the ratio of ionization fractions.  The term inside brackets of eq.\ (17) is typically close to unity 
(within a factor of two).  We scale to local-cloud temperatures $T = (7000~{\rm K})T_{7000}$, 
with constants $a_{\rm H}$ and $a_{\rm He}$ defined as 
\begin{eqnarray}
   a_{\rm H}  &=& \frac {\Gamma_{\rm HI} }  { n_{\rm H} \, \alpha_{\rm H} }  
               = (0.147) \left[ \frac {\Gamma_{\rm HI} } {10^{-14}~{\rm s}^{-1} } \right] 
                   \left[ \frac { n_{\rm H} } { 0.2~{\rm cm}^{-3} } \right] ^{-1} \, T_{7000}^{0.809}   \; ,  \\
   a_{\rm He} &=& \frac {\Gamma_{\rm HeI} } { n_{\rm H} \, \alpha_{\rm He} }   
    = (0.139) \left[  \frac {\Gamma_{\rm HeI} } {10^{-14}~{\rm s}^{-1} } \right] 
          \left[ \frac { n_{\rm H} } { 0.2~{\rm cm}^{-3} } \right] ^{-1} \, T_{7000}^{0.789} \; .
 \end{eqnarray}
The ionization ratio $y/x$ changes with depth inside the clouds, as changes in the shape of the  
spectra alter the ionization rates, $\Gamma_{\rm HI}$ and $\Gamma_{\rm HeI}$.  Because of the 
rapid decrease in photoionization cross sections at shorter wavelengths, the He-ionizing photons 
($\lambda \leq 504$~\AA) are more penetrating than those nearer to the \HI\ edge (700--900~\AA).  
For example, if $\Gamma_{\rm HI} = \Gamma_{\rm HeI} = 10^{-14}~{\rm s}^{-1}$ and 
$n_{\rm H} = 0.2~{\rm cm}^{-3}$, we find $x = 0.305$, $y = 0.294$, and $f_e = 0.334$.
If we increase $\Gamma_{\rm HeI}$ to $3\times10^{-14}~{\rm s}^{-1}$ we obtain 
($x, y) = (0.296, 0.544)$ and $f_e = 0.35$ at the cloud surface. Thus, the observed elevated 
He$^+$ fractions, $x =$ 0.2--0.3, $y =$ 0.4--0.5 and $f_e =$ 0.24--0.35, require a higher \HeI\
ionization rate $\Gamma_{\rm HeI} \approx$ (2--3)~$\Gamma_{\rm HI}$. 

\subsection{EUV Emission from the Local Hot Bubble} 

\medskip

Repeated SN explosions spread over several Myr have produced a produced a large cavity of hot, 
low-density plasma known as the Local Hot Bubble (LHB).  The energetics and timing of this bubble 
remain uncertain, but its spatial extent and physical properties have been observed in soft X-rays by 
ROSAT (S. L.\ Snowden \etal\ 1990; R. K.\ Smith \etal\ 2014), eROSITA (M. C. H.\ Yeung \etal\ 2024), 
and sounding rockets (e.g., D.\ McCammon \etal\ 2002;  M.\ Galeazzi \etal\ 2014).  An estimated 
10--20 SNe may have occurred over the last 10--12~Myr in the sub-groups of the Sco-Cen OB association 
(J.\ Ma\'iz-Apell\'aniz 2001; D.\ Breitschwerdt \etal\ 1996; B.\ Fuchs \etal\ 2006;  C.\ Zucker \etal\ 2025).  
D.\ Breitschwerdt \etal\ (2016) suggested that the most recent SNe may have occurred 1.5--2.6 Myr ago 
in the Upper Centaurus Lupus and Lower Centaurus Crux subgroups, if these events can be connected 
to $^{60}$Fe found in deep-sea crusts and estimated nucleosynthetic yields of core-collapse SNe.    

\medskip

In addition to the stellar sources, we have explored models that include the EUV radiation emanating 
from hot plasma  in the LHB.  The importance of this emission was previously modeled by 
J. D.\ Slavin \& P. C.\ Frisch (2002, 2008), where the harder EUV spectrum could explain the elevated 
He$^+$/H$^+$ ionization ratios.  At $T \approx 10^6$~K, most of the EUV is produced by electron-impact 
excited line emission from high ions of heavy elements, primarily Fe, Ne, Mg, and Si.   A complex of 
strong emission lines at 170--185~\AA\ comes from ($3d \rightarrow 3p$) lines of iron ions 
(\FeVIII, \FeIX, \FeX). The LHB contribution to LIC photoionization is sensitive to plasma emissivity for 
temperatures $5.8 \leq \log T \leq 6.2$.  In addition, the weak 170--175~\AA\ Fe lines seen in a rocket 
spectrum of the soft-X-ray diffuse background (D.\ McCammon \etal\ 2002) prompted suggestions that 
the gas-phase Fe abundance may be reduced by a factor of 3--7, owing to depletion into dust grains 
inside the hot bubble.  Despite the high LHB temperatures, the grain sputtering lifetime at $10^6$~K
can be quite long at low proton densities, with 
$t_{\rm sp} \approx (100~{\rm Myr}) (n_p / 0.01~{\rm cm}^{-3})^{-1}$,  However, it is uncertain whether 
these rocket observations apply to the low-density LHB  plasma in the local bubble. The target area of 
the rocket observations included bright $1/4$--keV regions in a 1-sr field centered at 
$\ell = 90^{\circ}$, $b = +60^{\circ}$, but avoiding Loop~I and the North Polar Spur. The spectral resolution 
was 9~eV, and the weak Fe lines at 69--72~eV lie at the edge of the detector, where the throughput was 
quite small. 

\medskip

As shown in {\bf Figure 5}, we find that a hot bubble with solar metal abundances could produce ionizing 
photon fluxes $\Phi_{\rm H} =$~7000--9000~cm$^{-2}$~s$^{-1}$, comparable to or greater than those of
the B stars.  In our models with metal depletion and reduced gas-phase abundances of Fe, Mg, and Si, 
we find fluxes $\Phi_{\rm H} =$~2000--4000~cm$^{-2}$~s$^{-1}$.  
Even with some depletion of refractory elements into dust grains, the EUV lines of neon ions 
(\NeVII\ and \NeVIII) remain strong.  Importantly, this emission includes flux in the \HeI\ continuum 
(100--504~\AA) from a range of ionization states of iron (Fe$^{+8}$ to Fe$^{+12}$) that could produce
elevated He$^+$ ionization fractions.  {\bf Figure 6} illustrates the flux attenuation and run of ionization
fractions with depth into the cloud for a constant-density cloud irradiated by a composite spectrum from 
all five hot stars and the LHB.  Because the shorter wavelengths in the \HeI\ continuum suffer less attenuation, 
the gas deeper into the cloud experiences relatively higher \HeI\ photoionization rates, resulting in elevated 
ionization fractions of He$^+$ ($y > x$) entering the Solar system.  

\medskip

 In {\bf Table 2} we summarize important parameters for the five stars: their ionizing photon fluxes
 $\Phi_{\rm H}$ and photoionization rates $\Gamma_{\rm H}$ and $\Gamma_{\rm He}$ at the local cloud
 surface, and the intervening column densities $N_{\rm HI}$ used for the flux attenuation.  
We also list values of the fluxes and ionization rates for six LHB models, spanning a range of temperatures
($\log T = 5.9, 6.0,6.1$) with two abundance sets (solar and depleted metals).  The EUV line emission 
from the LHB is comparable to that from the stars, and much of it occurs at wavelengths below the \HeI\ 
edge ($\lambda \leq 504.26$~\AA). 

\bigskip

\subsection {Stellar Motions and Ionization History}  

\bigskip

In the low density ISM, He$^+$ and H$^+$ may fall out of photoionization equilibrium, particularly
as a result of the motion of $\epsilon$~CMa and $\beta$~CMa relative to the Sun over the past 4.4~Myr.
For $\beta$~CMa, Hipparcos measurements (F.~van Leeuwen 2007) provide a parallax distance 
$d = 151\pm5$~pc,  radial velocity $V_r = 33.7\pm0.5$~\kms, and transverse velocity 
$V_{\perp} = 2.38\pm0.21$~\kms.  The transverse velocity is found from its proper motion 
$\mu_{\perp} = -3.32\pm0.28~{\rm mas~yr}^{-1}$ with individual components in right ascension 
(RA) and declination ($\delta$) given by $\mu_{\alpha} \,\cos \delta  = -3.23\pm0.19~{\rm mas~yr}^{-1}$ 
and $\mu_{\delta} = -0.78\pm0.20~{\rm mas~yr}^{-1}$.  Thus, $\beta$~CMa passed by the Sun 
at a time $t_{\rm pass} = d/V_r = 4.38\pm0.16$~Myr ago, at an offset distance 
$d_{\perp} = \mu_{\perp} d^2/V_r = 10.6\pm1.1$~pc.  These values are similar to those for
$\epsilon$~CMa, which passed by the Sun $4.44\pm0.10$~Myr at offset distance 
$9.3\pm0.5$~pc (J. M.\ Shull \etal\ 2025).   At that time, these two stars would have produced 
100--200 times higher ionization rates on any gas in the Sun's vicinity.   However, the local 
clouds are currently moving transverse to the Sun's vector at mean velocity $16.8\pm4.6$~\kms\
(P. C.\ Frisch \etal\ 2011).  Because the local clouds are gravitationally unbound from the Sun, their
peculiar motions and hydrodynamical forces from LHB gas make their past distance from the Sun 
difficult to predict.  Another consideration is that the B-giants, $\epsilon$~CMa and $\beta$~CMa, 
were closer to the main sequence 4~Myr ago, and thus fainter and hotter. The larger $T_{\rm eff}$ 
would offset the lower luminosity, keeping them important EUV sources.  

\medskip

Over the past 4--5 Myr, $\epsilon$~CMa and $\beta$~CMa had strong ionizing effects on nearby 
interstellar gas, including a wake of ionized gas along their past trajectories -- the interstellar tunnel 
of highly ionized gas discussed in the next subsection.
Several papers (J. L.\ Linsky \& S. Redfield  2021; C.\ Zucker \etal\ 2025) explored the possibility that 
the tunnel and local-cloud complex were created by ionization fronts and Str\"omgren spheres 
around nearby hot stars.  The radial velocities of the B-stars correspond to motions of 27.9~pc/Myr 
($\epsilon$~CMa) and 34.5~pc/Myr ($\beta$~CMa) away from the Sun.  Tracking the ionization 
rates $\Gamma_{\rm H}$ along their past trajectories, we find that the ionization fraction inside
warm, diffuse clouds ($T = 7000$~K, $n_{\rm H} = 0.2~{\rm cm}^{-3}$) within 10~pc of the B-stars 
would rise to $x \approx 0.99$.   After this ``flash ionization", non-equilibrium ionization effects may 
set in, as the recombination time is longer than the photoionization time at the cloud surface.
As listed below, the cooling time adopts a radiative cooling rate per volume $n_{\rm H}^2 \Lambda(T)$ 
with coefficient $\Lambda(T) \approx 3\times10^{-26}~{\rm erg~cm}^3~{\rm s}^{-1}$ at $T \approx 7000$~K.  
This cooling rate would increase (and $t_{\rm cool}$ would decrease) as the clouds are heated above 
$10^4$~K, owing to excitation of \Lya\ and metal lines,
\begin{eqnarray}
    t_{\rm ph}    &=& \Gamma_{\rm HI}^{-1} \approx (1~{\rm Myr}) 
                              \left( \frac {\Gamma_{\rm HI}}{3\times10^{-14}~{\rm s}^{-1}}  \right)^{-1} \\
    t_{\rm rec}  &=& (n_e \alpha_{\rm H})^{-1}  \approx (0.93~{\rm Myr}) 
                             \left( \frac {n_e}{0.1~{\rm cm}^{-3}} \right)^{-1} T_{7000}^{0.809}  \\
    t_{\rm cool} &=&  \frac {3n_{\rm tot} kT/2} { n_{\rm H}^2 \, \Lambda(T)} 
                          \approx (11~{\rm Myr}) \left( \frac {n_{\rm H}}{0.2~{\rm cm}^{-3}} \right)^{-1}  
                           T_{7000}    \; . 
\end{eqnarray}
Here, we used a case-B recombination rate coefficient  
$\alpha_{\rm H} = (3.39\times10^{-13}~{\rm cm}^3~{\rm s}^{-1})T_{7000}^{-0.809}$ (B.\ Draine 2011) at
$T = (7000~{\rm K})T_{7000}$,  total hydrogen density $n_{\rm H} \approx 0.2$~cm$^{-3}$, and electron
density $n_e \approx 0.1~{\rm cm}^{-3}$.  We scaled $\Gamma_{\rm HI} $ to the (current) combined 
ionization rates for $\beta$~CMa and $\epsilon$~CMa, which dominate the stellar photoionizing continua. 
These rates would rise by a factor of 100 or more 4.4~Myr ago.  

\bigskip

\subsection{Local Tunnel Shape and Ionization}

\bigskip

We now discuss the origin of the tunnel of low N$_{\rm HI}$ in the direction ($\ell = 232^{\circ}\pm5^{\circ}$, 
$b = -15^{\circ}\pm4^{\circ})$ toward the two B-stars at Galactic coordinates ($\ell$, $b$) of ($239.83,-11.33$) 
for $\epsilon$~CMa and ($226.06,-14.27)$ for $\beta$~CMa.  
This cavity may have been formed by nearby Str\"omgren spheres (J. L.\ Linsky \& S.\ Redfield 2021) 
or shock waves and gas flows produced by SNe in the Sco-Cen OB association (P. C.\ Frisch \etal\ 2011; 
M.\ Piecka et al. 2024).  The local clouds could have been compressed by D-critical shock waves 
preceding ionization fronts (L.\ Spitzer 1978).   Recently, C.\ Zucker \etal\ (2025) addressed whether 
the clouds were influenced by ionization fronts associated with Str\"omgren spheres or supernova 
explosions within a pre-evacuated LHB.   They concluded that SNe were primarily responsible for 
the structure of the clouds over the past Myr, as their remnants propagated into a low-density medium 
with $n_{\rm H} = 0.04~{\rm cm}^{-3}$.

\medskip 

In analyzing these possibilities, stellar timing is crucial.  The hot, low-density cavity likely already 
existed when the two B-stars passed by the Sun.  The gas in the local tunnel would then be quite 
hot ($10^6$~K), having been swept by gas flows from the Sco-Cen SNe.  In this case, the 
Str\"omgren-sphere model is not an accurate description of the physical conditions.  The evidence 
for partial ionization of H and He in the local clouds is the result of the low flux of ionizing radiation,
not the edge of the Str\"omgren zone.  In the usual formulation for the Str\"omgren sphere, applied 
to $\epsilon$~CMa and $\beta$~CMa  with $n_{\rm H} = 0.04~{\rm cm}^{-3}$, the radius of the
ionized zone would be
 \begin{equation}
   R_s = \left[ \frac {3 Q_{\rm H}}{4 \pi n_e n_{\rm H} \alpha_{\rm H}(T)} \right]^{1/3} 
       \approx (56~{\rm pc}) Q_{46}^{1/3} 
       \left( \frac {\alpha_{\rm H}}{2.59\times10^{-13}~{\rm cm}^3~{\rm s}^{-1} } \right)^{-1/3}
       \left( \frac {n_{\rm H}} {0.04~{\rm cm}^{-3}}\right)^{-2/3}   \; .
 \end{equation}
 Here, we scaled to an ionizing photon luminosity 
 $Q_{\rm H} = (10^{46}~{\rm phot~s}^{-1})Q_{46}$ appropriate for the two CMa B-stars.
 To account for diffuse LyC photons from recombination to the ($n = 1$) ground state
 of \HI, we adopted a case-B recombination rate coefficient 
 $\alpha_{\rm H} = 2.59\times10^{-13}~{\rm cm}^{3}~{\rm s}^{-1}$ at $T = 10^4$~K.  However, 
 the tunnel plasma may be shock-heated to $T \approx 10^6$~K, similar to the LHB. 
In that case, the recombination rate coefficient is much lower, 
$\alpha_{\rm H} = 1.21 \times 10^{-14}~{\rm cm}^{3}~{\rm s}^{-1}$, and the LyC photons 
have extremely long mean-free paths (100~pc or more) in the tunnel.   

\medskip

The two B-stars have relatively low ionizing  fluxes compared to those in the \HII\  regions 
around massive O-type stars.  There should be no sharp boundaries in the ionized gas, as 
seen in the \HII\ regions around O-type stars.  Instead, the gas approaching the boundary is 
only partially ionized, and the Str\"omgren model is inappropriate.  Furthermore, in the hot 
tunnel gas, the collisional ionization rate of hydrogen greatly exceeds the combined 
photoionization rate of the two B-stars, with a ratio
 \begin{equation}
     \frac {n_e \, C_{\rm ion}} {\Gamma_{\rm H}} =
        \frac {1.4\times10^{-10}~{\rm s}^{-1}} {5.6\times10^{-12}~{\rm s}^{-1}}  \approx 25   \; .
 \end{equation} 
We adopted $\Gamma_{\rm H} = 5.6\times10^{-12}~{\rm s}^{-1}$, scaling the photoionization 
rates of $\epsilon$~CMa and $\beta$~CMa  at their current distances (124~pc and 151~pc) 
to a closer distance of 10 pc.  The rate coefficient
$C_{\rm ion} = 2.74\times10^{-8}~{\rm cm}^3~{\rm s}^{-1}$ for \HI\ collisional ionization at 
$10^6$~K was derived from integrated cross sections in  T. T.\ Scholz \& H. R. J.\ Walters (1991). 
In the hot gas in the tunnel interior, the neutral fraction would be quite small,
 \begin{equation}
    \frac {n_{\rm HI}} {n_{\rm H}} = \frac {\alpha_{\rm H}(T)}  {C_{\rm ion}(T)} =
        \frac {1.21 \times10 ^{-14}~{\rm cm}^{3}~{\rm s}^{-1}} 
       {2.74\times10^{-8}~{\rm cm}^{3}~{\rm s}^{-1}}  \approx 4.4\times10^{-7}   \; .
\end{equation}
The \HI\ column density over a 100-pc distance through the low-density tunnel, with 
$n_{\rm H} = 0.04~{\rm cm}^{-3}$, would only be $N_{\rm HI} = 5\times10^{12}~{\rm cm}^{-2}$.  
Instead of a reaching a Str\"omgren-type boundary within the local tunnel, the ionizing
photons will propagate freely over long distances through the tunnel toward the diffuse clouds 
around the Sun.   The LyC photons radiating transverse to the motion of the B-stars,
with flux $\Phi_{\rm H} \approx 8\times10^5$~cm$^{-2}$~s$^{-1}$ will be absorbed at the 
tunnel walls, with an ionization front moving at speed 
$V_I = \Phi_{\rm H}/n_{\rm H} \approx 8$~\kms\ for gas at density 
$n_{\rm H} \approx 1~{\rm cm}^{-3}$ at offset distance $\sim10$~pc consistent with 
the $\pm5^{\circ}$ angular extent of the tunnel.

\medskip

 We conclude that the local interstellar tunnel was initially shaped and heated by gas flows 
 and SNR blast waves from Sco-Cen.  When the two B-stars passed through this low-density 
 gas, their photoionization added to the collisional ionization in the tunnel interior.  
 There is some evidence (C. G.\ Gry \etal\ 1985) that $\beta$~CMa has now moved into a 
 denser \HII\ region, with a higher column density  of ionized hydrogen inferred from \SII\ 
 and \SiIII\ absorption. Because $\beta$~CMa (151~pc) is currently farther away than
 $\epsilon$~CMa (124~pc) and separated by $13.74^{\circ}$ on the sky, this defines the 
 radial and angular extent of the tunnel in those directions.            
                                                                                                   

 \newpage

\section{\bf Summary of Results}  

\bigskip

We derived a consistent set of stellar parameters for $\beta$~CMa (mass, radius, effective 
temperature, luminosity) consistent with its shorter parallax distance (151~pc vs.\ 206~pc), 
interferometric angular diameter ($\theta_d = 0.52\pm0.03$~mas), and integrated bolometric 
flux, $f = (36.2 \pm 4.9) \times10^{-6}~{\rm erg~cm}^{-2}~{\rm s}^{-1}$.  From these, we derived
$T_{\rm eff} = 25,180\pm1120$~K and updated values of its absolute magnitude 
($M_V = -3.93\pm0.04$ and $M_{\rm bol} = -5.97$).  

\medskip

\noindent
 The following points summarize our primary results:
\begin{enumerate}

\item The stellar parallax distance and angular diameter of $\beta$~CMa give a radius 
$R = 8.44\pm0.56~R_{\odot}$.  Bolometric relations between the integrated stellar flux and 
radius yield an effective temperature $T_{\rm eff} \approx 25,180\pm1120$~K and luminosity 
$L \approx 10^{4.41\pm0.06}~L_{\odot}$.  Both gravitational and evolutionary masses are 
consistent at $M \approx 13\pm1~M_{\odot}$.  

\item From models of the stellar atmosphere and interstellar attenuation of the ionizing flux 
in the Lyman continuum, we determine a column density 
$N_{\rm HI} = (1.9\pm0.1) \times 10^{18}~{\rm cm}^{-2}$ in the local cloud toward 
$\beta$~CMa, corresponding to optical depth $\tau_{\rm LL} = 12.0\pm0.6$ at the Lyman limit.   

\item Using non-LTE model atmospheres and observed EUV spectra, we estimate a stellar
flux decrement $\Delta_{\rm star} = 67\pm7$ at the Lyman limit of $\beta$~CMa. 
At the external surface of the local cloud, its ionizing photon fluxes (cm$^{-2}$~s$^{-1}$) are
$\Phi_{\rm HI} \approx 3700\pm1000$ and $\Phi_{\rm HeI} \approx 110 \pm 30$ in the
\HI\ and \HeI\ continua.   From its distance $d = 151\pm5$~pc, we find a LyC  photon 
production rate  $Q_{\rm H} \approx 10^{46.0\pm0.1}~{\rm photons~s}^{-1}$. 
 
\item  Because of attenuation, the EUV flux from $\beta$~CMa incident on the local clouds 
is $\sim20$ times higher than viewed from Earth.  The photoionization rates at the cloud
surface are $\Gamma_{\rm HI} \approx 1.5 \times 10^{-14}~{\rm s}^{-1}$ and 
$\Gamma_{\rm HeI} \approx 7.3\times10^{-16}$~s$^{-1}$.  For local gas with 
$n_{\rm H} = 0.2~{\rm cm}^{-3}$ and $T \approx 7000$~K, photoionized solely by
$\epsilon$~CMa and $\beta$~CMa, the ionization fractions at the cloud surface would
be $x \approx 0.48$ (H$^+$) but only $y \approx 0.02$ (He$^+$).  They would have been 
much higher in the past (4.4 Myr ago) when both $\epsilon$~CMa and $\beta$~CMa passed 
within $10\pm1$~pc of the Sun.

\item The elevated He$^+$ ionization fractions in the local cloud ($y$ = 0.4--0.5) compared 
to H$^+$ ($x$ = 0.2--0.3), with free electron fractions $f_e = (x+0.1y) \approx$ 0.24--0.35, 
require $\Gamma_{\rm HeI} \approx$ (2--3)$\Gamma_{\rm HI}$ in photoionization equilibrium.    
The EUV emission from Fe, Ne, Mg  ions in million-degree local bubble plasma produces
an ionizing photon flux $\Phi_{\rm HI} \approx$~7000--9000~cm$^{-2}~{\rm s}^{-1}$ (at solar 
abundances), and its harder EUV spectrum may explain the He$^+$/H$^+$ ionization ratios.    
However, He$^+$ and H$^+$ could be out of photoionization equilibrium, after a period of
high ionization 4.4 Myr ago, when both $\epsilon$~CMa and $\beta$~CMa passed near the
Sun.  At that time, the local clouds were farther from the Sun, owing to their transverse motion.  
However, the two CMa B-stars left a wake of hot (photoionized and collisionally ionized) gas in 
the interstellar tunnel, with extremely long LyC photon mean free paths of 100 pc, comparable
 to LHB dimensions.  

\end{enumerate} 

The Sun is currently in a special location, surrounded by a hot cavity of million-degree gas and 
shielded from EUV emission by \HI\ and \HeI\ in diffuse clouds within the local 10~pc.  In the future, 
the Sun will exit these clouds and once again be exposed to the higher ionizing radiation from the
B-stars, white dwarfs, and hot bubble.  


\begin{acknowledgements}

 We thank Emily Witt, James Green, and Kevin France for discussions of the far-UV 
 spectra of $\beta$~CMa with the Colorado DEUCE rocket and Dan McCammon
 for discussions on the soft-X-ray background observations.  We thank Cecile Gry, 
 Jeffrey Linsky, Seth Redfield, John Vallerga, Barry Welsh, and Pawel Swaczyna  for 
 discussions about the local ISM.  We appreciate the referee's prompt report and the
 suggestion to expand the scope to include other ionizing sources of the local ISM.  
 A portion of this study was supported by the {\it New Horizons Mission} program for 
 astrophysical studies of cosmic optical, ultraviolet, and \Lya\ backgrounds.

\end{acknowledgements}

\newpage


       
 \newpage


\appendix

 \section{\bf  Cloud Ionization Fractions and Densities}
 
 \medskip
 
 Measurements of \HI\ and \HeI\ absorption by EUVE in the local clouds toward white 
 dwarfs (J.\ Dupuis \etal\ 1995) suggested excess ionization of He$^+$ relative to H$^+$. 
The most direct observations of the neutral densities of \HeI\ (He$^0$) and \HI\ (H$^0$) 
in the local cloud were made by the {\it Voyager}, {\it Ulysses}, and {\it IBEX} spacecraft
(M.\ Witte 2004; G.\ Gloeckler \etal\ 2009;  E.\ M\"obius \etal\ 2015).
The inferred densities of He$^0$ and H$^0$ inflow to the heliosphere were based on 
impact measurements of pickup ions, corrected for filtration due to charge exchange in 
the solar wind (P.\ Swaczyna \etal\ 2023).  These effects are severe for H, but less so for He.  
Here, we derive analytic formulae for estimates of the ionization fractions
($x$, $y$) of hydrogen and helium and the total number density $n_{\rm H}$ of hydrogen.  
These are based on EUVE observations of the neutral abundance ratio 
$N_{\rm HI} /N_{\rm HeI}$, together with densities of interstellar electrons ($n_e$) from 
{\it Hubble Space Telescope} absorption lines and inflowing neutral helium ($n_{\rm HeI}$)
 from the {\it Ulysses} and {\it IBEX} spacecraft.

\medskip
 
We want to determine three model parameters ($x$, $y$, $n_{\rm H}$) based on four 
observed quantities for the local cloud, denoted ($A$, $\beta$, $n_e$, $n_{\rm HeI}$).  
Below, we derive formulae for $x$, $y$, and $n_{\rm H}$ in terms of $A$, $\beta$, and 
the density ratio $R \equiv (n_e / n_{\rm HeI})$.  These observational parameters include
$A = n_{\rm He}/n_{\rm H}$, the solar He abundance (M.\ Asplund \etal\ 2021), and 
$\beta = N_{\rm HI}/N_{\rm HeI}$, the ratio of neutral column densities of H and He in 
the local interstellar cloud.   From EUVE spectra of six nearby hot white dwarfs,
 J.\ Dupuis \etal\ (1995) reported an elevated  interstellar \HI/\HeI\ ratio $\beta \approx 14$, 
 after correcting the observed \HI\ and \HeI\ column densities for stellar contributions using 
 LTE, hydrostatic, plane-parallel model atmospheres.  The EUVE  column densities showed
 $\beta$ ranging from $12.1\pm3.6$ to $18.4\pm1.3$.  Removing the two extreme values 
 (the lowest value of 12.1 has 30\% uncertainty) gives a mean and rms dispersion of  
 $\beta = 14.5\pm1.1$.  Most of these sight lines pass through more than the local cloud, 
 making $\beta$ a somewhat uncertain measure of the interstellar \HeI/\HI\ outside the 
 heliosphere.  Nevertheless, we adopt it as typical of the LIC.

\medskip

A second observational parameter is the number density of neutral helium ($n_{\rm HeI}$) 
in the local cloud as it flows into the heliosphere.  Observations with instruments on the
{\it Ulysses} and {\it Voyager} spacecraft found $n_{\rm HeI} \approx 0.015~{\rm cm}^{-3}$ 
(G.\ Gloeckler \etal\ 2009) with errors quoted variously as $\pm0.0015~{\rm cm}^{-3}$ 
(G.\ Gloeckler \& J.\ Geiss 2004) and $\pm 0.002~{\rm cm}^{-3}$  (G.\ Gloeckler \& J.\ Geiss 
2001; G.\ Gloeckler \etal\ 2009).  The direct measurements of He$^+$ pickup ionls were corrected 
for filtration and charge-exchange in the solar wind, adding some systematic model uncertainty.  
If we multiply $n_{\rm HeI} = 0.015\pm0.002~{\rm cm}^{-3}$ by the mean EUVE ratio 
$N_{\rm HI}/N_{\rm HeI} = 14.5\pm1.1$, we can estimate the density of interstellar neutral hydrogen, 
$n_{\rm HI} \approx 0.22\pm0.03~{\rm cm}^{-3}$, entering the solar wind at the termination shock. 
The total hydrogen density would be somewhat higher, after including ionized hydrogen.  
P.\ Swaczyna \etal\ (2022) noted that the inflowing gas appears to have a higher density than 
sight lines in other directions.  They suggested that the elevated density comes from a 
``mixing region" produced by interaction between the LIC and G-cloud. 

\medskip

The third observational parameter is the electron density in the local clouds.   Previous 
estimates for the LIC found $n_e  = 0.12 \pm 0.05~{\rm cm}^{-3}$ in the direction of 
$\epsilon$~CMa (C. G.\ Gry \& E. B.\ Jenkins 2001) and 
$n_e  = 0.11^{+0.025}_{-0.03}~{\rm cm}^{-3}$ toward $\alpha$~Leo 
(C. G.\ Gry \& E. B.\ Jenkins 2017).  The electron density was found from modeling
interstellar column-density ratios (\MgII/\MgI) and collisionally excited fine-structure 
populations in the ground state of \CII.
S.\ Redfield \& R.~E.\ Falcon (2008) modeled $n_e$ for multiple sight lines, finding a 
log-normal distribution with an unweighted mean of 
$n_e =0.11^{+0.10}_{-0.05}~{\rm cm}^{-3}$.  Other reviews quoted similar densities, 
including $n_e = 0.10\pm0.04~{\rm cm}^{-3}$ 
(P.\ Frisch \etal\ 2011) and $n_e = 0.07\pm0.01~{\rm cm}^{-3}$ 
(J. D.\ Slavin \& P. C.\ Frisch 2008).  In the analytic model below, we adopt values of 
$n_e = 0.10\pm0.03~{\rm cm}^{-3}$ and $n_{\rm HeI} = 0.015\pm0.002~{\rm cm}^{-3}$.
 
 \medskip

We begin with key relations among model parameters ($x$, $y$, $n_{\rm H}$) and 
assumptions about He/H and other observational constraints.  We adopt abundance ratios
$A = n_{\rm He} / n_{\rm H} = 0.1$ and $\beta = n_{\rm HI} / n_{\rm HeI} = 14.5 \pm 1.1$.  
We assume homogeneous ionization conditions, with (\HI/\HeI) number density ratios 
tracking the EUVE column densities and electron density $n_e = n_{\rm H} (x+Ay)$.
This is a rough approximation, since the radiation field and ionization fractions ($x,y$)
change with depth into the cloud.  However, it allows us to make estimates of cloud 
ionization fractions, using two expressions involving the neutral helium fraction 
$n_{\rm HeI}/n_{\rm He} \equiv (1-y)$.  The first relation is
\begin{equation}
     \frac {n_{\rm HeI}} {n_{\rm HI} } = \frac {n_{\rm He} (1-y)}{n_{\rm H} (1-x)} \; \; \; 
          {\rm or} \; \; \;     (1-y) = \frac {(1-x)} { A \, \beta }      \; .
 \end{equation}
The definitions $n_{\rm HeI} = A n_{\rm H}(1-y)$  and $n_e = n_{\rm H} (x+Ay)$,
lead to the second relation,
 \begin{equation}
    (1-y) = \frac {n_{\rm HeI}} {A \, n_{\rm H}} = \frac {n_{\rm HeI} (x+Ay)} {A\,n_e}  \;  .
 \end{equation}
 We equate the two expressions for $(1-y)$ to find 
 \begin{equation}
   \beta \, n_{\rm HeI} \, (x + Ay) = n_e (1-x)    \;  .
 \end{equation}
To simplify the expressions, we denote the key density ratio  
\begin{equation} 
   R \equiv  \frac {n_e}{n_{\rm HeI}}  = 
       \frac { (0.10 \pm 0.03~{\rm cm}^{-3}) } {(0.015\pm0.002~{\rm cm}^{-3})} = 6.67\pm2.19 \; .
\end{equation} 
Substituting for $y = 1 - [(1-x)/A\beta]$, we solve for $(1-y) = [(1+A)/A (R + \beta + 1)]$,
allowing us to express the ionization fractions of H and He as
\begin{equation} 
   x  = \frac { [ R - A \, \beta +1] } { [R + \beta + 1] } = 0.280  \; \; \;  \;  {\rm and} \; \; \; 
   y =  \frac { [(R + \beta) - (1/A)] }  { [R + \beta + 1] } = 0.504  \; . 
\end{equation}
The total hydrogen density of the cloud, $n_{\rm H} = [n_{\rm HeI} / A \, (1-y)]$,
is then found to be 
 \begin{equation}
    n_{\rm H} = \frac {n_{\rm HeI}} {(1+A)} \left[  R + \beta + 1 \right]  
                         \approx 0.302~{\rm cm}^{-3} \; .
\end{equation}
The numerical values above assume $A = 0.1$, $\beta = 14.5\pm1.1$, and $R = 6.67\pm2.19$.
Considering the variations in $\beta$ alone, we find ranges in ionization of $x =$ 0.26--0.30 and 
$y =$ 0.48--0.52.  For variations in just $R$, we find $x =$ 0.21--0.34 and $y =$ 0.45--0.55.  
Considering uncertainties in both $\beta$ and the density ratio $R$, we find a range of model 
parameters,
\begin{equation}
x = 0.28^{+0.06}_{-0.07}  \; \; \; \;  y = 0.50^{+0.04}_{-0.05}  \; \; \; \;
n_{\rm H} = 0.30^{+0.03}_{-0.02}~{\rm cm}^{-3}    \; . 
\end{equation}
More accurate values require photoionization models that account for the attenuation 
of ionizing radiation with depth into the cloud, as both  $\Gamma_{\rm H}$ and 
$\Gamma_{\rm He}$ decrease from the cloud external surface inward toward 
the heliosphere.
 
 \bigskip
 
            
 
 
 \section{\bf Coupled Photoionization of Hydrogen and Helium}  

\bigskip

We model the LIC in photoionization equilibrium, in which the ionization states of 
hydrogen and helium balance the rates of photoionization and radiative recombination,
\begin{eqnarray}
   n_{\rm HI}  \,  \Gamma_{\rm HI}   &=& n_e \, n_{\rm HII} \,  \alpha_{\rm H}(T) \; ,   \\
   n_{\rm HeI} \, \Gamma_{\rm HeI} &=& n_e \, n_{\rm HeII} \,  \alpha_{\rm He}(T)   \; .
\end{eqnarray}
We assume that neutral hydrogen and neutral helium are photoionized at rates (s$^{-1}$)
\begin{eqnarray} 
   \Gamma_{\rm HI} &=& \int_{\nu_{\rm H}}^{\infty} \frac {F_{\nu}}{h \nu}  \; \sigma_{\rm HI} (\nu) \, d \nu \; ,  \\
   \Gamma_{\rm HeI} &=& \int_{\nu_{\rm He}}^{\infty} \frac {F_{\nu}}{h \nu} \; \sigma_{\rm HeI} (\nu) \, d \nu  \; . 
\end{eqnarray}
We integrate the frequency distribution of energy flux $F_{\nu}$ and photoionization cross sections
$\sigma(\nu)$ in the ionizing continua of neutral hydrogen 
($h \nu  \geq 13.5984$~eV, $\lambda \leq 911.75$~\AA) and neutral helium
($h \nu \geq 24.5874$~eV, $\lambda \leq 504.26$~\AA).  The ions (H$^+$ and He$^+$)
are neutralized by radiative recombination with rate coefficients $\alpha_{\rm H}$ and
$\alpha_{\rm He}$. We neglect the doubly ionized state of helium (He$^{+2}$), which can be 
shown to be quite small (less than 1\%) in the local clouds.

\medskip

We define the ionization fractions of H and He as $x = n_{\rm HII}/n_{\rm H}$ and 
$y = n_{\rm HeII} / n_{\rm He}$, where $n_{\rm H}$ and $n_{\rm He}$ are the total number 
densities of H and He including both neutrals and ions.  The electron number density is
$n_e = n_{\rm HII} + n_{\rm HeII}$, which we express in fractional form 
$f_e =  n_e / n_{\rm H} = x+0.1y$ using $n_{\rm He} / n_{\rm H} = 0.1$.  
The photoionization equilibrium equations can then be written
\begin{equation}
     \frac {x}{(1-x)} =  \frac { a_{\rm H} }   {f_e} \; \; \; {\rm and} \; \; \;  
     \frac {y}{(1-y)} =  \frac { a_{\rm He} } {f_e}        \; , 
\end{equation} 
where the constants $a_{\rm H}$ and $a_{\rm He}$ are defined as 
\begin{eqnarray}
   a_{\rm H}  &=& \frac {\Gamma_{\rm HI} }  { n_{\rm H} \, \alpha_{\rm H} }  
               = (0.147) \left[ \frac {\Gamma_{\rm HI} } {10^{-14}~{\rm s}^{-1} } \right] 
                   \left[ \frac { n_{\rm H} } { 0.2~{\rm cm}^{-3} } \right] ^{-1} \, T_{7000}^{0.809}   \; ,  \\
   a_{\rm He} &=& \frac {\Gamma_{\rm HeI} } { n_{\rm H} \, \alpha_{\rm He} }   
    = (0.139) \left[  \frac {\Gamma_{\rm HeI} } {10^{-14}~{\rm s}^{-1} } \right] 
          \left[ \frac { n_{\rm H} } { 0.2~{\rm cm}^{-3} } \right] ^{-1} \, T_{7000}^{0.789} \; .
 \end{eqnarray}
 The case-B recombination rate coefficients to \HI\ and \HeI\ are approximated as
\begin{eqnarray}
   \alpha_{\rm H }(T)   &=& (3.39\times10^{-13}~{\rm cm}^3~{\rm s}^{-1} ) T_{7000}^{-0.809} \; ,  \\
   \alpha_{\rm He} (T) &=& (3.60\times10^{-13}~{\rm cm}^3~{\rm s}^{-1} ) T_{7000}^{-0.789}  \; , 
\end{eqnarray} 
for temperatures near the fiducial LIC value $T = (7000~{\rm K})T_{7000}$.  
The electron fraction $f_e = (x+0.1y$) is determined consistently (with $x$ and $y$) 
by iteration from the coupled relations
\begin{equation}
   x = \frac { (a_{\rm H}/f_e) }   {[1 + (a_{\rm H}/f_e) ] }   \; \; \; {\rm and}   \; \; \; 
   y = \frac { (a_{\rm He}/f_e) } {[1 + (a_{\rm He}/f_e) ]  }  \;  .
\end{equation} 
These provide a useful relation for the ratio of He-to-H ionization fractions,
\begin{equation}
   \frac  {y}{x} = \left( \frac {\Gamma_{\rm He}}  {\Gamma_{\rm HI}} \right) 
         \left[ \frac {1 + (a_{\rm H}/f_e)} {1 + (a_{\rm He}/f_e)} \right] T_{7000}^{-0.020}  \;  .
\end{equation}


\clearpage 


\begin{deluxetable} {lccc cccc}
\tablecolumns{8}
 \tabletypesize{\scriptsize}

\tablenum{1}
\tablewidth{0pt}
\tablecaption{Various Stellar ($\beta$~CMa) Parameters\tablenotemark{a} }  

\tablehead{
   \colhead{Reference Paper}
 & \colhead{$d$}
 & \colhead{$T_{\rm eff}$} 
 & \colhead{$\log g$} 
 & \colhead{$R/R_{\odot}$}
 & \colhead{$M/M_{\odot}$} 
 & \colhead{$L/L_{\odot}$} 
 & \colhead{$M_{\rm bol}$}
 \\
 \colhead{}
 & \colhead{(pc)}
 & \colhead{(K)} 
 & \colhead{(cgs)} 
 & \colhead{}
 & \colhead{} 
 & \colhead{} 
 & \colhead{(mag)} 
 }

\startdata
Cassinelli \etal\ (1996)  & 206 & $25,180\pm1130$ & $3.4\pm0.15$ & $16.2^{+1.2}_{-1.2}$  
                        & $15.2^{+6.4}_{-4.4}$ & $45,900\pm9500$ & $-6.91$  \\
Fossati \etal\ (2015)\tablenotemark{b} & $151\pm5$ & $24,700\pm300$ & $3.78\pm0.08$ & $7.4^{+0.8}_{-0.9}$ 
                         & $12.0^{+0.3}_{-0.7}$ & $25,700^{+3800}_{-3800}$ & $-6.29$ \\ 
Fossati \etal\ (2015)\tablenotemark{c}  & $151\pm5$ & $24,700\pm300$ & $3.78\pm0.08$ & $8.2^{+0.6}_{-0.5}$ 
                         & $12.6^{+0.4}_{-0.5}$ & $25,700^{+3800}_{-3300}$ & $-6.29$  \\ 
Current Study (2025) & $151\pm5$  & $25,180\pm1120$ & $3.70\pm0.08$ & $8.44\pm0.56$ & $13\pm1$ 
                         & $25,800\pm3900$  & $-5.97$    \\
\enddata 

\tablenotetext{a} {Values of effective temperature, surface gravity, radius, mass, luminosity,
and bolometric absolute magnitude given in past papers.  L.\ Fossati \etal\ (2015) derived
$R$ and $M$ from two sets of evolutionary tracks (footnotes b and c). }

\tablenotetext{b} {Stellar mass and radius inferred from evolutionary tracks of C.\ Georgy \etal\ (2013). }
\tablenotetext{c} {Stellar mass and radius inferred from evolutionary tracks of L.\ Brott \etal\ (2011). }

\end{deluxetable}



\begin{deluxetable}  {llccclll}
    \tablecolumns{8}
    \tabletypesize{\footnotesize}
\tablenum{2}
\tablewidth{0pt}
\tablecaption{Photoionization Rates\tablenotemark{a} }  

\tablehead{
 \colhead{Ionizing Source}
 &  \colhead{Type} 
 &  \colhead{ $T_{\rm eff}$ }
 & \colhead{Distance} 
 & \colhead{$\Phi_{\rm H}$}
 & \colhead{$\Gamma_{\rm HI}$}
 & \colhead{$\Gamma_{\rm HeI}$} 
 & \colhead {$N_{\rm HI}$} 
 \\
   \colhead{  }
 & \colhead{  } 
 & \colhead {(K)} 
 & \colhead{(pc)} 
 & \colhead{(cm$^{-2}$~s$^{-1}$)}
 & \colhead{(s$^{-1}$)} 
 & \colhead{(s$^{-1}$)} 
 & \colhead{(cm$^{-2}$)} 
 }

\startdata
{\bf Stellar Sources:}   &                          &              &                        &            &              &               &            \\
$\epsilon$~CMa          &  B1.5~II            & 21,000  & $124\pm 2$    & 3100   & 1.4E-14 & 6.2E-17 & 6.0E17  \\
$\beta$~CMa              &  B1~II-III           & 25,180   & $151\pm 5$    & 3700   & 1.5E-14 & 1.0E-15 & 1.9E18  \\
Feige~24                     &  WD (DA1)       & 53,000   & $77.5\pm0.2$ &   870   & 2.2E-15 & 2.9E-15 & 2.9E18  \\
G191-B2B                   &  WD  (DA0)      & 57,000   & $52.5\pm0.2$ &   490   & 1.1E-15 & 1.7E-15 & 1.5E18  \\
HZ~43A                       &  WD (DA1)       & 50,400   & $60.3\pm0.2$ &  180    & 4.3E-16 & 5.4E-16 & 8.5E17  \\
                                    &                          &              &                        &            &              &               &               \\
{\bf Local Hot Bubble:} &                          &              &                        &            &              &               &             \\
Solar Abundances       & $\log T = 5.9$   &  \dots     &  out~to~85     & 8700   & 9.8E-15 & 3.9E-14 & 1.0E18  \\
Solar Abundances       & $\log T = 6.0$   &  \dots     &  out~to~85     & 7800   & 5.8E-15 & 3.1E-14 & 1.0E18  \\
Solar Abundances       & $\log T = 6.1$   &   \dots    &  out~to~85     & 7000   & 3.8E-15 & 2.6E-14 & 1.0E18  \\
                                    &                           &              &                        &            &              &               &               \\
Depleted Abundances & $\log T = 5.9$   &   \dots    & out~to~85      & 4000   & 8.6E-15 & 1.6E-14 & 1.0E18  \\
Depleted Abundances & $\log T = 6.0$   &   \dots    & out~to~85      & 2300   & 3.9E-15 & 8.4E-15 & 1.0E18  \\
Depleted Abundances & $\log T = 6.1$   &   \dots    & out~to~85      & 1800   & 2.2E-15 & 6.5E-15 & 1.0E18  \\
\enddata 

\tablenotetext{a} {Sources of ionizing photons incident on local interstellar clouds include two B-type 
stars,  three white dwarfs, and the Local Hot Bubble.  Parallax distances are from {\it Hipparcos} 
(B-stars) and {\it Gaia} (WDs).  We also list stellar effective temperatures $T_{\rm eff}$ and integrated 
ionizing photon fluxes $\Phi_{\rm H}$  and photoionization rates  $\Gamma_{\rm HI}$ and $\Gamma_{\rm HeI}$ 
at the cloud surface in continua of H~I ($\lambda \leq 912$~\AA) and He~I ($\lambda \leq 504$~\AA).  
Values for the Local Hot Bubble were calculated using the {\it Chianti} code for plasma at three temperatures. 
We adopted solar abundances including [Fe/H] $= 6.46 \pm 0.04$ (M.\ Asplund \etal\ 2021) and constant 
electron density $n_e = 0.004~{\rm cm}^{-3}$ out to bubble radius $R = 85$~pc from the Sun. If refractory
elements (Fe, Si, Mg) are depleted by a factor of 5, the bubble fluxes and ionization rates drop with 
reduced Fe/H gas-phase abundances, but EUV lines from the Ne ions remain strong (see Fig.\ 5).  }
 
\end{deluxetable}



\begin{figure}[ht]
\centering
\includegraphics[angle=0,scale=0.9] {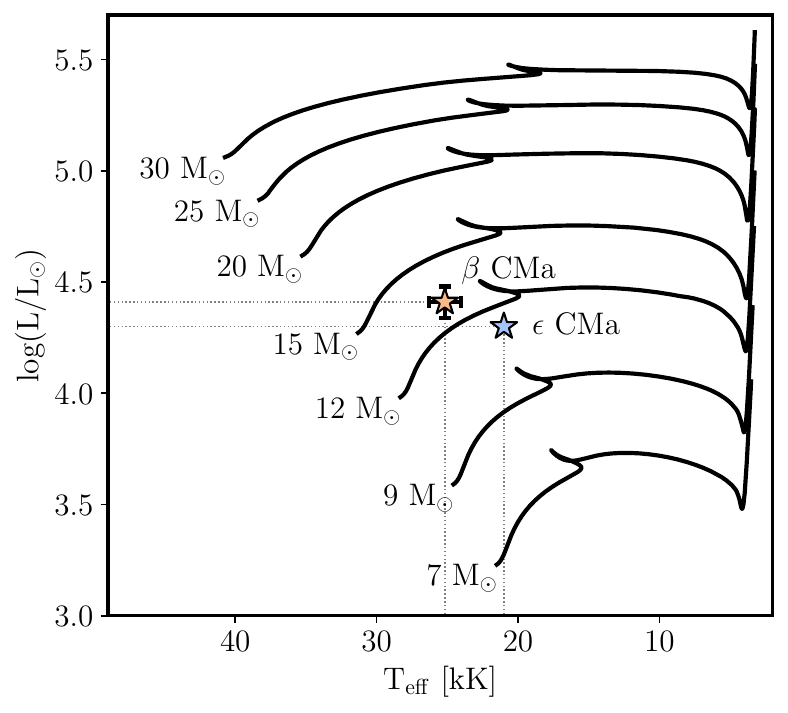}
\caption{The location of $\beta$~CMa on the Hertzsprung-Russell diagram is shown for 
our derived parameters, $\log (L/L_{\odot}) = 4.41\pm0.06$ and 
$T_{\rm eff} = 25,180\pm1120$~K, based on new radius $R = 8.44\pm0.56~R_{\odot}$ 
and parallax distance $d = 151\pm5$~pc.  The evolutionary tracks are from L.\ Brott \etal\ (2011) 
with Milky Way metallicities and initial masses labeled from 7--30 $M_{\odot}$.  
The location of $\epsilon$~CMa (J. M.\ Shull \etal\ 2025) is shown for comparison.
 }
\end{figure}


\begin{figure}[ht]
\includegraphics[angle=0,scale=0.57] {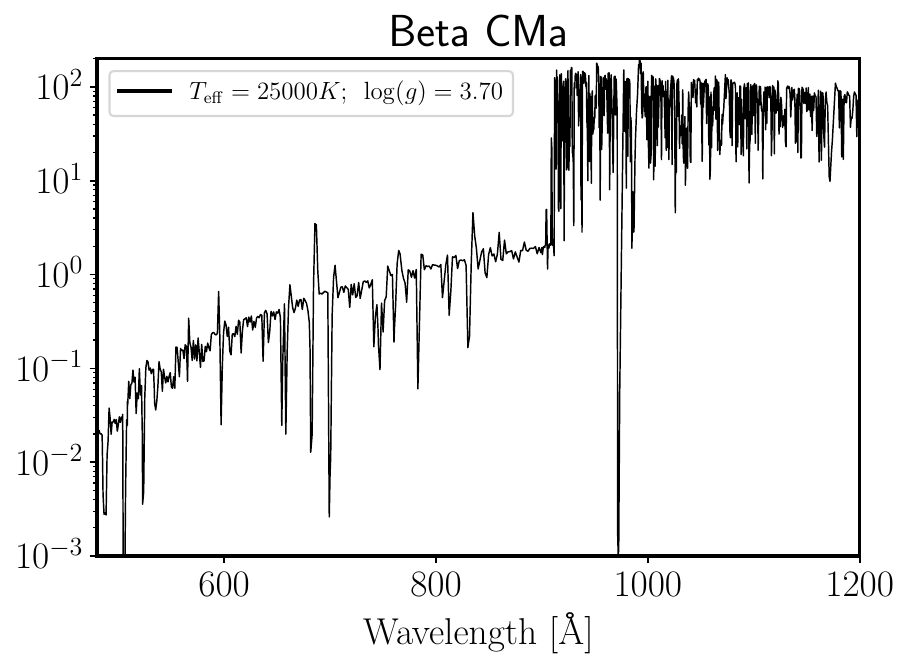}
\includegraphics[angle=0,scale=0.57] {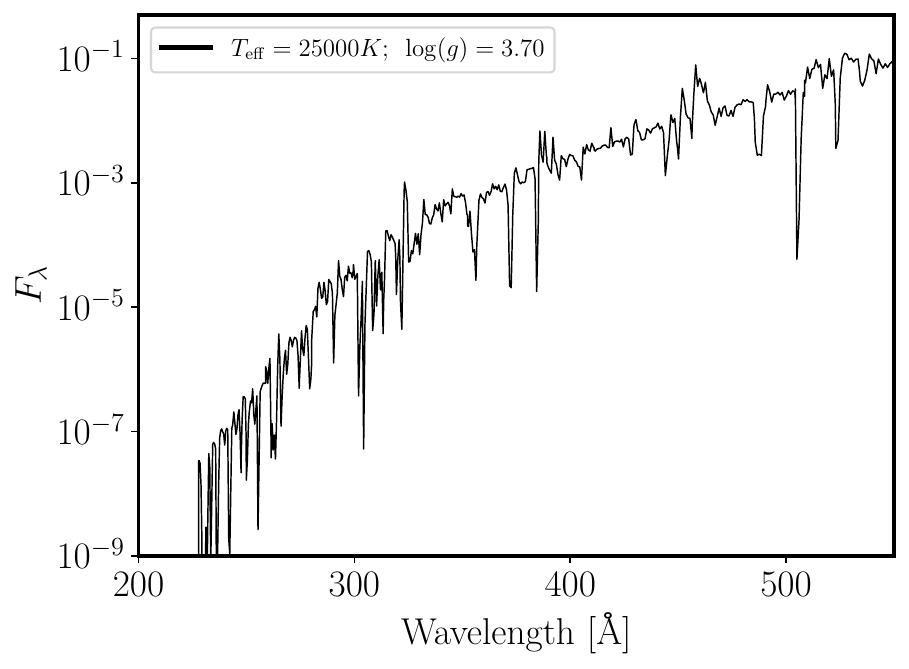}
\caption{Far-UV and EUV spectra for $\beta$~CMa from a model atmosphere
computed with the non-LTE line-blanketed code \texttt{WM-basic} for effective 
temperature $T_{\rm eff} = 25,000$~K  and surface gravity $\log g = 3.70$. 
(Left)   Flux distribution $\log F_{\lambda}$ from 500--1200~\AA, showing the 
Lyman limit decrement at 912~\AA.  
(Right)  Flux distribution from 228~\AA\ to 550~\AA.  The absence of an edge
at the \HeI\ ionization limit (504~\AA) is a result of non-LTE effects from 
backwarming of the upper atmosphere from a wind in early B-type stars.  
}

\end{figure}



\begin{figure}[ht]
\centering
\includegraphics[angle=0,scale=0.50] {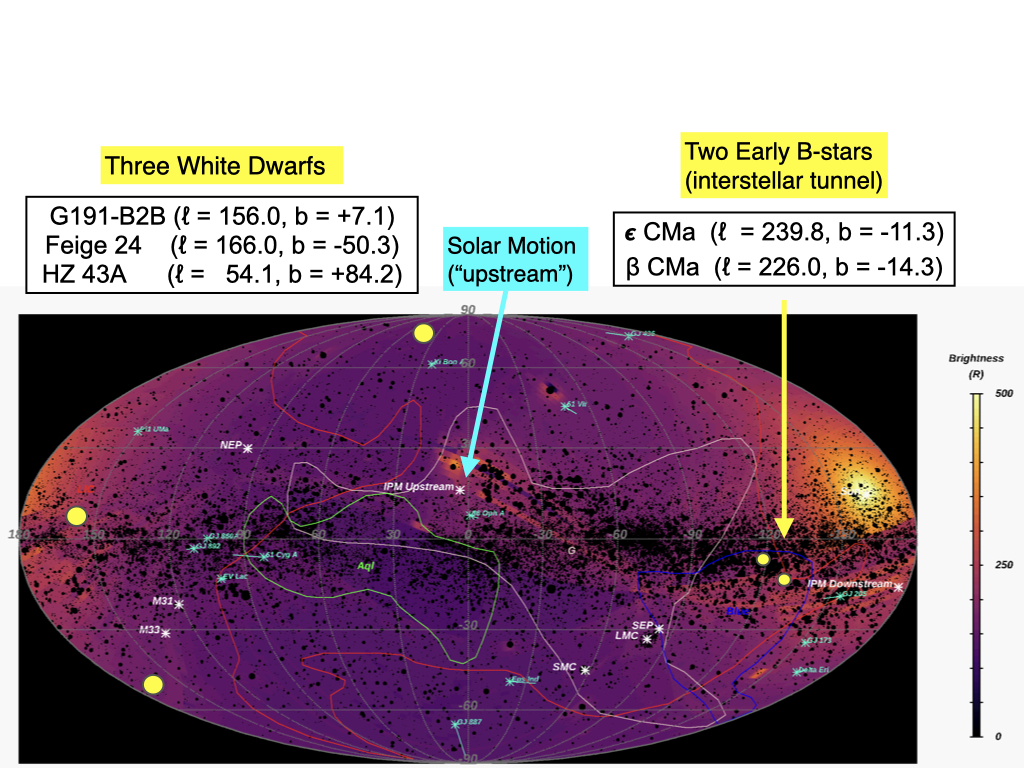}
\caption{Locations of the five stars (yellow circles) that dominate the ionization of the local clouds,
shown in Galactic coordinates ($\ell$, $b$) centered on (0, 0).   Locations are plotted on the 
All-sky New Horizons Alice \Lya\ map (G. R.\ Gladstone \etal\ 2025) taken at 57 AU from the Sun.  
Wavy lines indicate the outlines of four of the important local interstellar clouds (LIC in red; Aql in green; 
Blue  in blue; and G in tan).  The map also indicates locations of the Sun (right edge), north and
south ecliptic poles (NEP, SEP), notable stars and galaxies, and the ``upstream and downstream"
directions of flow of interstellar \HI\ through the solar system and interplanetary medium (IPM).  
Black dots show the $\sim90,000$ stars in the M. A.\ Velez \etal\ (2024) catalog of potential sources 
of far-UV emission.  
 }
\end{figure}


\begin{figure}[ht]
\centering
\includegraphics[angle=0,scale=0.75] {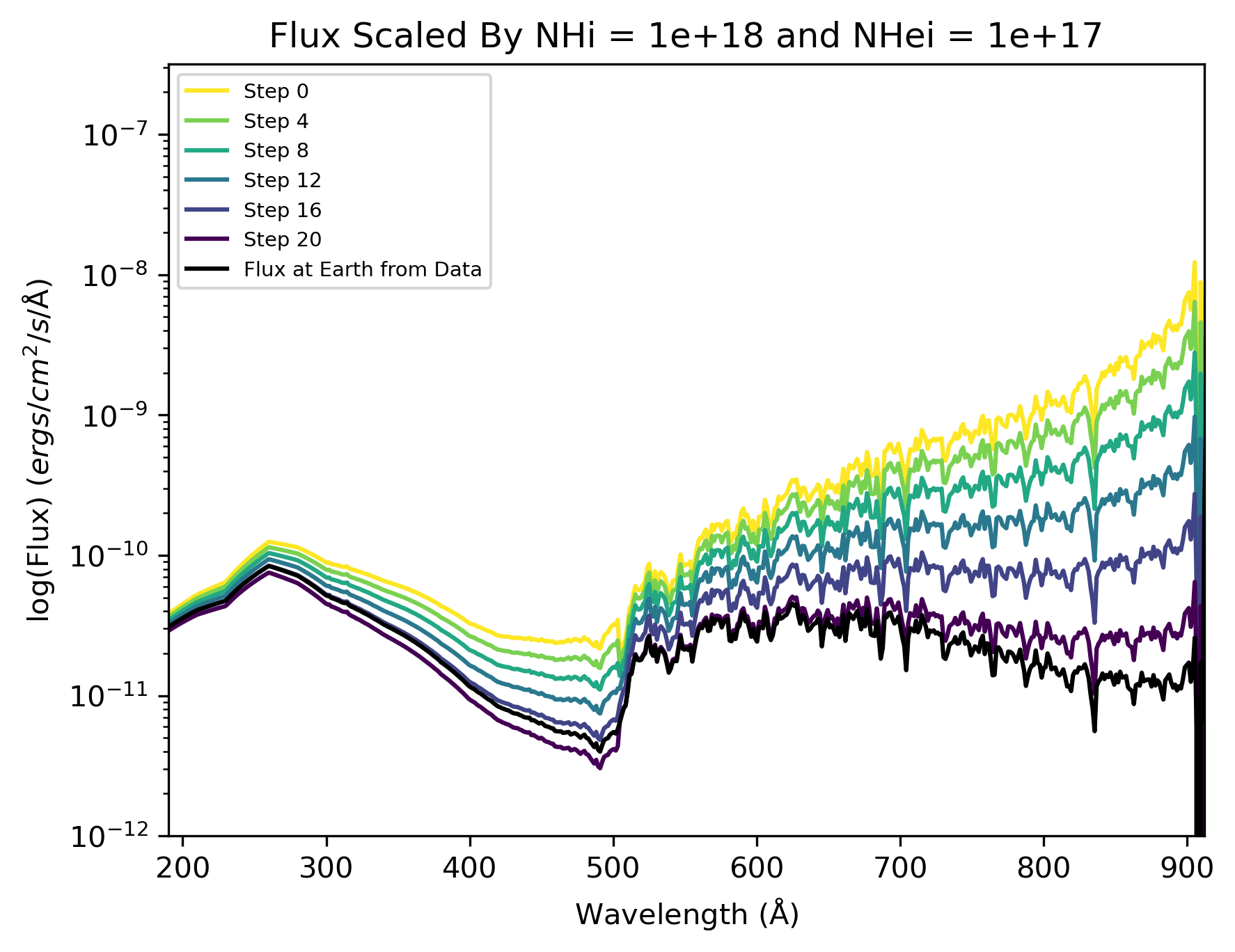}
\includegraphics[angle=0,scale=0.75] {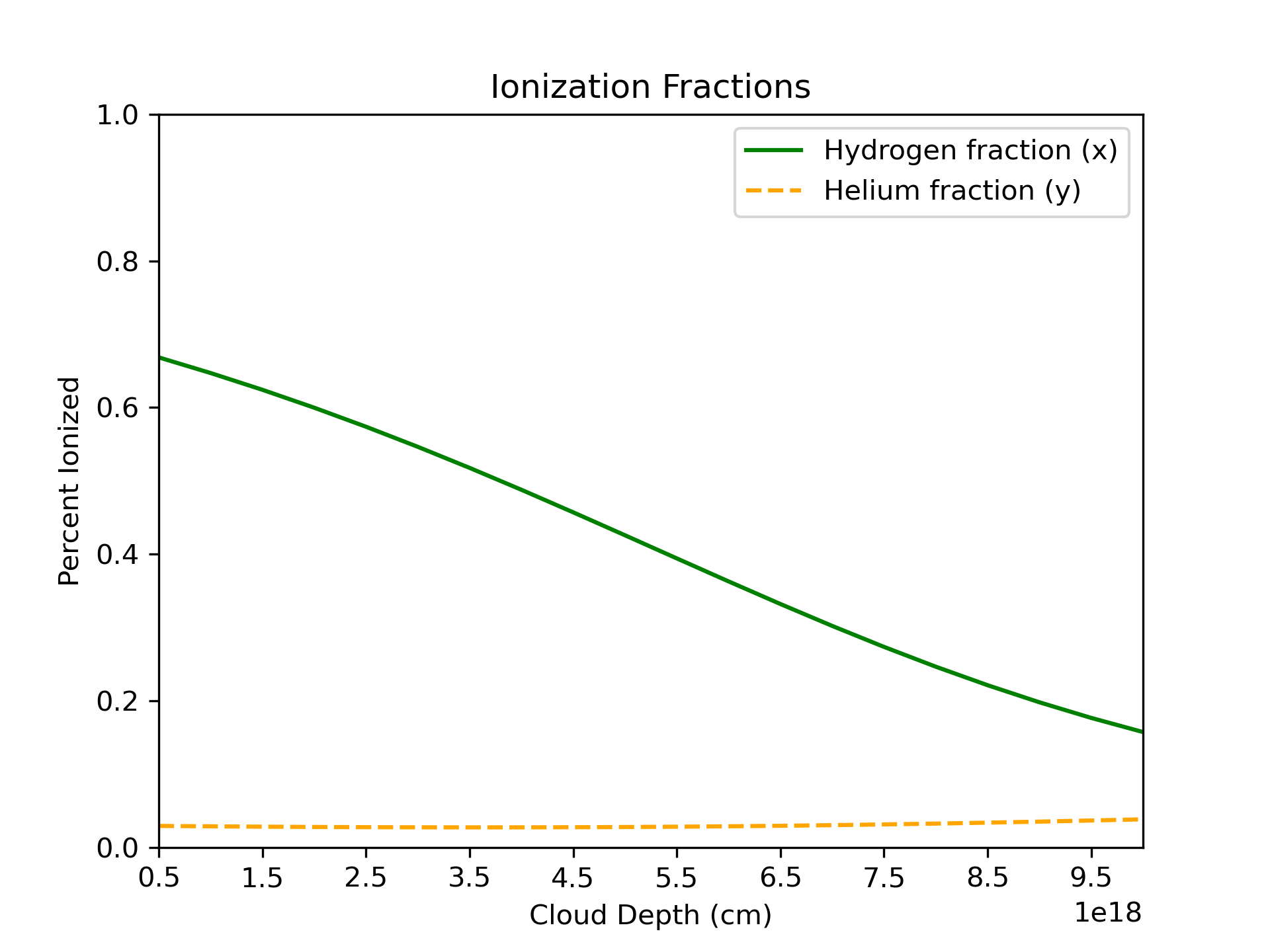}

\caption{Model of constant density cloud ($n_{\rm H} = 0.2$~cm$^{-3}$, $T = 7000$~K) 
using ionizing fluxes from all five stellar sources (two B-stars, three white dwarfs) but
not the hot bubble.  Fluxes are attenuated by mean column densities 
$N_{\rm HI} = 10^{18}~{\rm cm}^{-2}$ and $N_{\rm HeI} = 10^{17}~{\rm cm}^{-2}$. 
Spectra and ionization fractions are shown at various depths into the cloud, with distance 
intervals from step-0 (external surface) to step-20 (entering heliosphere) and step length 
$\Delta L = 5\times10^{17}$~cm.  
(Top)  Attenuated spectra with depth into the local cloud, with the bottom spectrum 
(black) showing the attenuated flux seen at Earth.  
(Bottom) Ionization fractions of H and He with depth. 
 }
\end{figure}


\begin{figure}[ht]
\centering
\includegraphics[angle=0,scale=0.45] {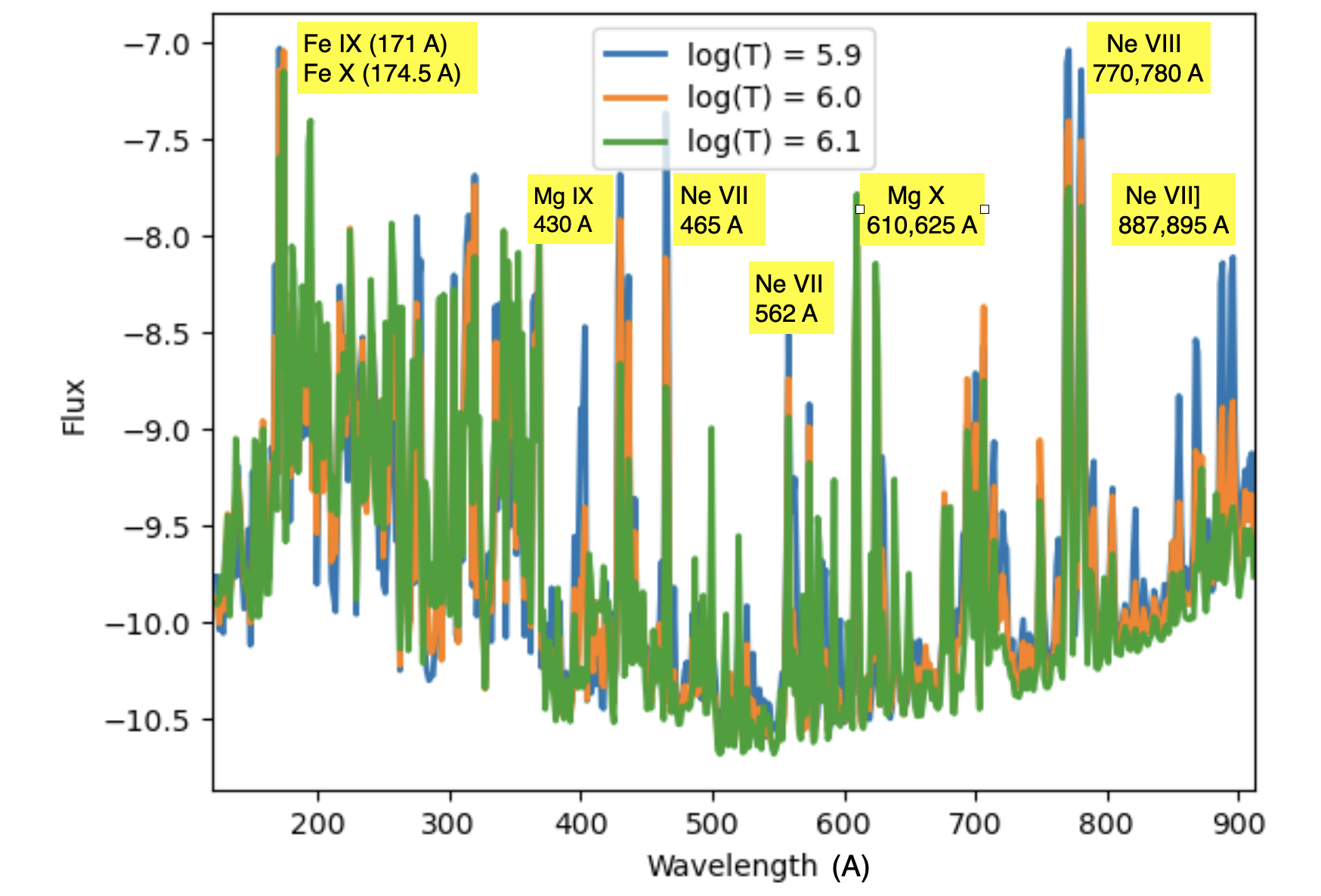}
\includegraphics[angle=0,scale=0.40] {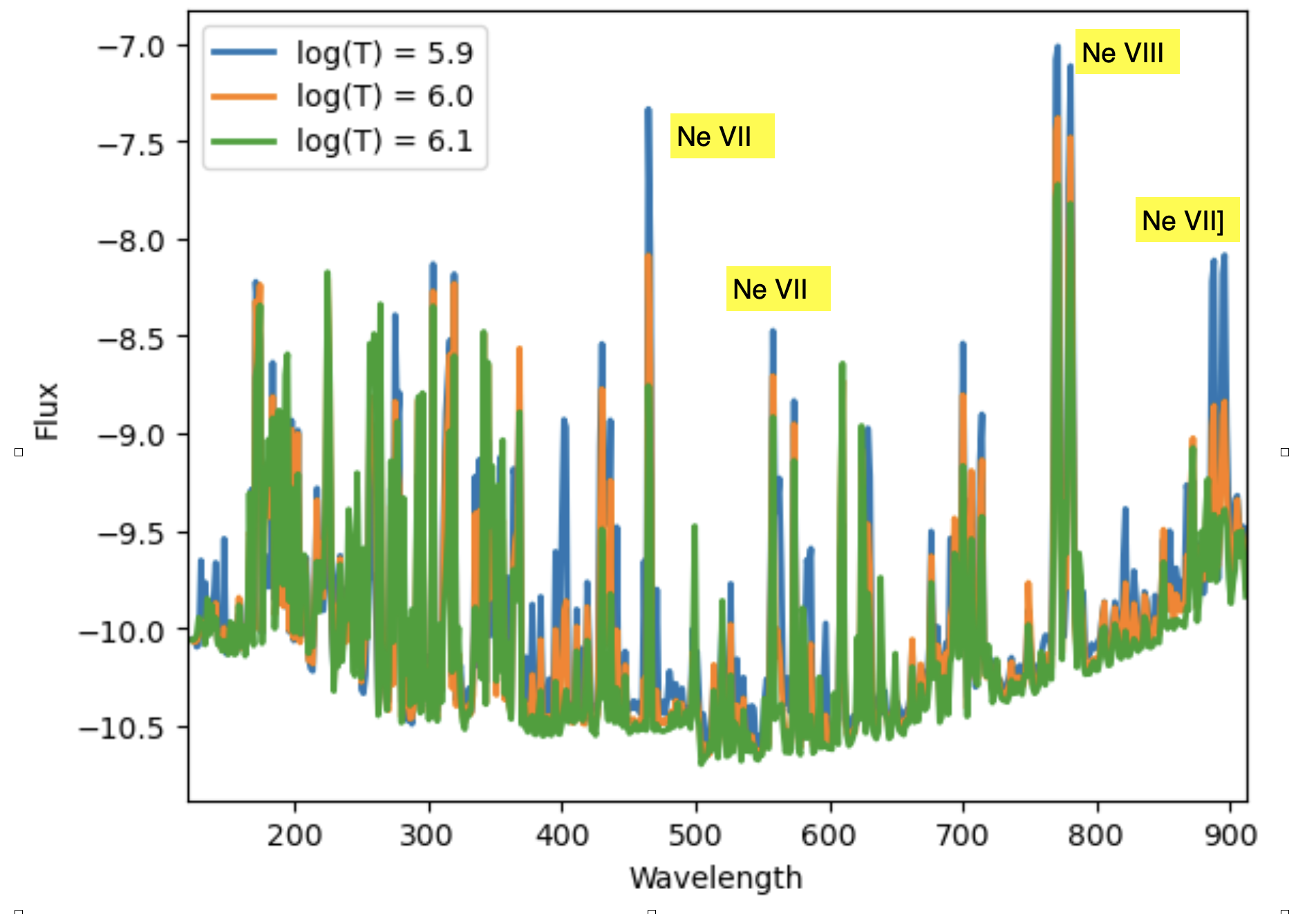}

\caption{Modeled spectral distribution of the flux, $\log F_{\lambda}$ 
(in erg~cm$^{-2}~{\rm s}^{-1}~{\rm \AA}^{-1}$), of ionizing (EUV) photons produced 
in the Local Hot Bubble, calculated using the {\it Chianti} code for plasma at three 
temperatures.  We assume a constant electron density $n_e = 0.004~{\rm cm}^{-3}$, 
with EUV emissivity integrated out to a bubble radius $R = 85$~pc from the Sun.  
The emissivities of Fe and Ne ions decline with temperature over the range 
$\log T = 5.9, 6.0, 6.1$, with color-coded fluxes (blue, orange, green) as labeled in 
box.  Prominent EUV emission lines are noted.
(Top panel) Fluxes with solar metal abundances including [Fe/H] $ = 6.46 \pm 0.04$ 
(M.\ Asplund \etal\ 2021).
(Bottom panel)  Fluxes with refractory elements (Fe, Mg, Si) reduced in abundance by a 
factor of 5, owing to depletion into dust grains.  Neon is not depleted, and its lines remain 
strong.  The underlying continuum is from bremsstrahlung and radiative recombination.   
 }
\end{figure}


\begin{figure}[ht]
\centering
\includegraphics[angle=0,scale=0.7] {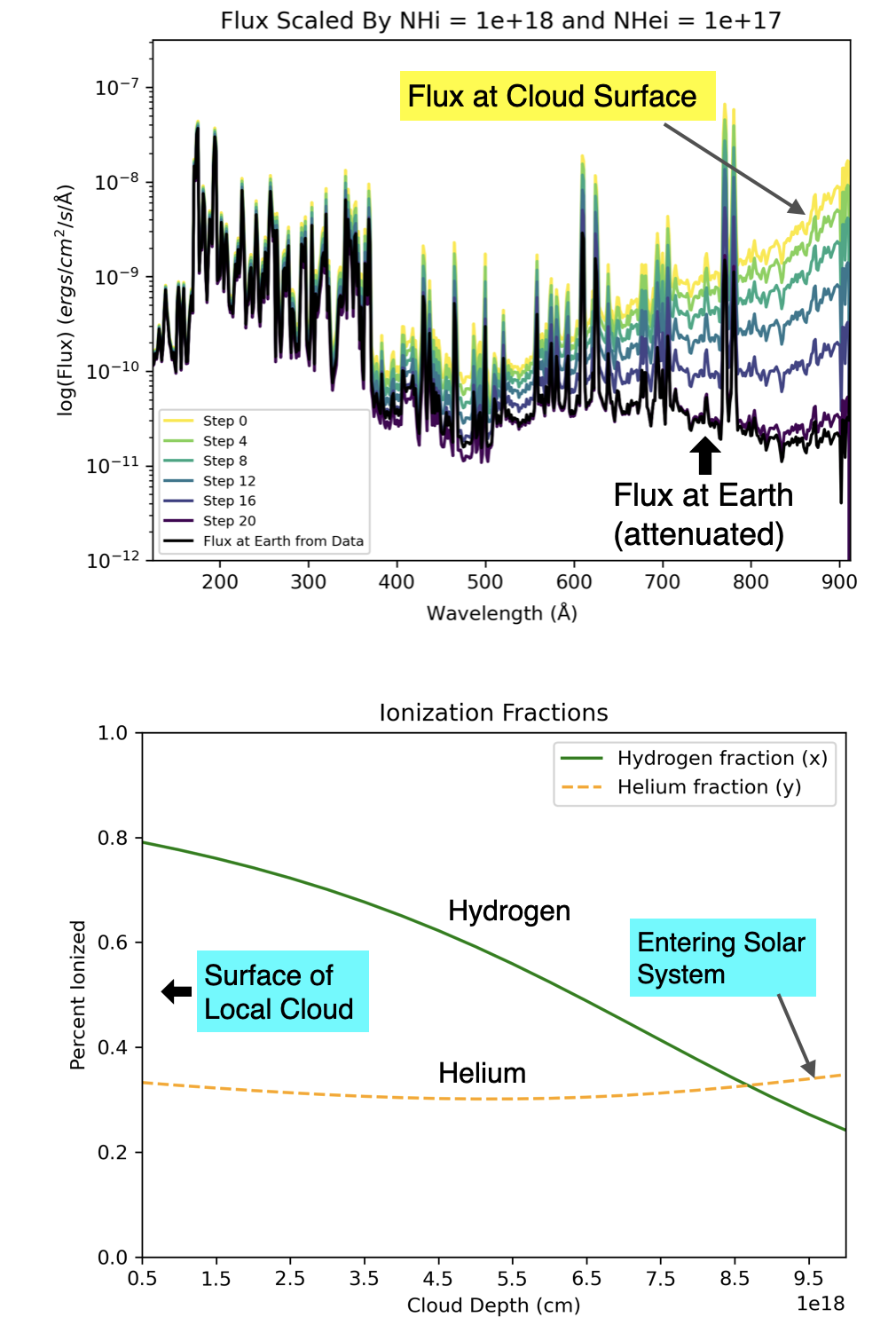}

\caption{Model of attenuated EUV spectrum and ionization fractions of H and He.  Parameters
are the same as in Fig.\ 4, but we now include the line emission from the LHB (solar abundances) 
along with the five stellar sources.  This produces elevated He$^+$ ionization fractions deeper
into the cloud.  Spectra and ionization are shown at various depths, at distance intervals from 
step-0 (external surface) to step-20 (entering heliosphere) and step length 
$\Delta L = 5\times10^{17}$~cm.  
(Top) Ionizing spectra at several distances into the cloud, from surface to Earth.
(Bottom) Ionization fractions of H and He with depth. }
\end{figure}

\end{document}